\documentclass[a4paper,twocolumn,10pt,accepted=2025-08-04]{quantumarticle}
\pdfoutput=1
\usepackage{amssymb}
\usepackage{color}
\usepackage{amsmath}
\usepackage{amsbsy}
\usepackage{amsthm}
\usepackage{graphicx}
\usepackage{bbm}
\usepackage{bm}
\usepackage{epsfig}
\usepackage{xfrac}
\usepackage{xcolor}
\usepackage{enumerate}
\usepackage{multirow}
\usepackage{physics}
\usepackage[T1]{fontenc}
\usepackage[english]{babel}
\usepackage{symbols}
\usepackage{makecell}
\usepackage{appendix}
\usepackage{dsfont}
\usepackage{float}
\usepackage{pdfpages}
\usepackage{placeins}
\usepackage{array}
\usepackage{relsize}
\usepackage{xcolor}
\usepackage{scalerel}
\usepackage{tikz}
\usetikzlibrary{svg.path}
\usepackage{hyperref}
\usepackage[numbers,sort&compress]{natbib}

\makeatletter
\AtBeginDocument{\let\LS@rot\@undefined}
\makeatother

\begin{document}

\title{Quantum control of continuous systems via nonharmonic potential modulation}

\author{Piotr~T.~Grochowski}
\email{piotr.grochowski@upol.cz}
\affiliation{\href{https://ror.org/056hzt889}{Institute for Quantum Optics and Quantum Information of the Austrian Academy of Sciences}, A-6020 Innsbruck, Austria}
\affiliation{Institute for Theoretical Physics, \href{https://ror.org/054pv6659}{University of Innsbruck}, A-6020 Innsbruck, Austria}
\affiliation{Department of Optics, \href{https://ror.org/04qxnmv42}{Palacký University}, 17. listopadu 1192/12, 771 46 Olomouc, Czech Republic}
\orcid{0000-0002-9654-4824}

\author{Hannes~Pichler} 
\affiliation{\href{https://ror.org/056hzt889}{Institute for Quantum Optics and Quantum Information of the Austrian Academy of Sciences}, A-6020 Innsbruck, Austria}
\affiliation{Institute for Theoretical Physics, \href{https://ror.org/054pv6659}{University of Innsbruck}, A-6020 Innsbruck, Austria}
\orcid{0000-0003-2144-536X}

\author{Cindy~A.~Regal} 
\affiliation{\href{https://ror.org/008hybe55}{JILA}, National Institute of Standards and Technology and University of Colorado, Boulder, Colorado 80309, USA}
\affiliation{Department of Physics, \href{https://ror.org/02ttsq026}{University of Colorado}, Boulder, Colorado 80309, USA}
\orcid{0000-0002-0000-2140}

\author{Oriol~Romero-Isart} 
\affiliation{\href{https://ror.org/056hzt889}{Institute for Quantum Optics and Quantum Information of the Austrian Academy of Sciences}, A-6020 Innsbruck, Austria}
\affiliation{Institute for Theoretical Physics, \href{https://ror.org/054pv6659}{University of Innsbruck}, A-6020 Innsbruck, Austria}
 \affiliation{\href{https://ror.org/03g5ew477}{ICFO - Institut de Ciencies Fotoniques}, The Barcelona Institute of Science and Technology, 08860 Castelldefels (Barcelona), Spain}
 \affiliation{\href{https://ror.org/0371hy230}{ICREA}, Passeig Lluis Companys 23, 08010 Barcelona, Spain}
 \orcid{0000-0003-4006-3391}

\begin{abstract}
We present a theoretical proposal for preparing and manipulating a state of a single continuous-variable degree of freedom confined to a nonharmonic potential.
By utilizing optimally controlled modulation of the potential's position and depth, we demonstrate the generation of non-Gaussian states, including Fock, Gottesman-Kitaev-Preskill, multi-legged-cat, and cubic-phase states, as well as the implementation of arbitrary unitaries within a selected two-level subspace.
Additionally, we propose protocols for single-shot orthogonal state discrimination, algorithmic cooling, and correcting for nonlinear evolution.
We analyze the robustness of this control scheme against noise.
Since all the presented protocols rely solely on the precise modulation of the effective nonharmonic potential landscape, they are relevant to several experiments with continuous-variable systems, including the motion of a single particle in an optical tweezer or lattice, or current in circuit quantum electrodynamics.
\end{abstract}

\maketitle

\section{Introduction}
\begin{figure}[H]
    \centering
    \includegraphics[width=\linewidth]{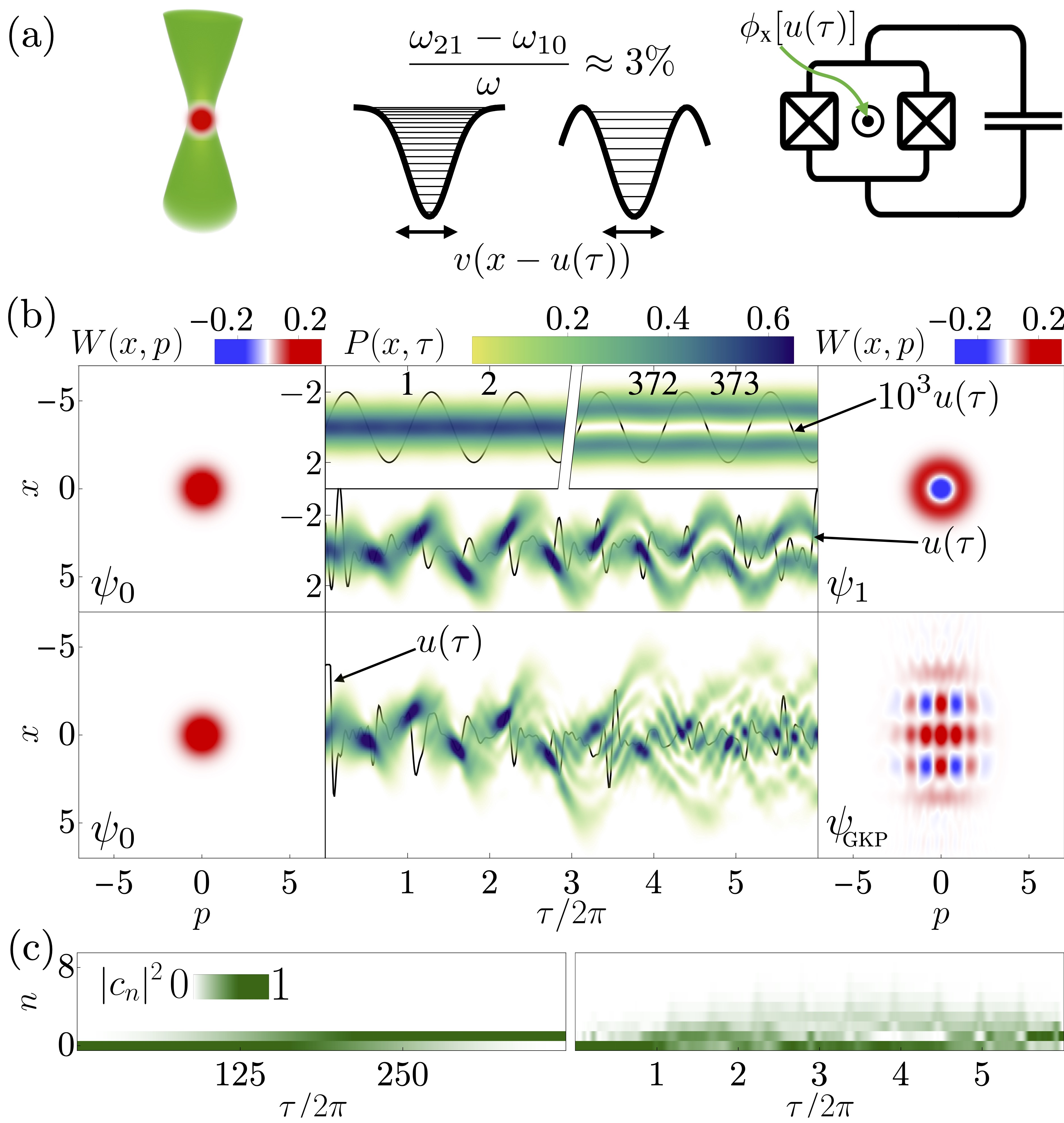}
    \caption{(a) Examples of continuous-variable nonlinear systems that can be optimally controlled without the need for auxiliary systems---single atoms in optical tweezers and flux-tunable transmons.
    (b) Time-evolved probability density $P(\Pos,\Time)=|\WaveFunction (\Pos,\Time)|^2$, during the state preparation protocol with the snapshots of Wigner functions $W(\Pos,\Mom)$ at the beginning and the end of the protocol.
    The potential is Gaussian and its optimally controlled position $\PosControl(\Time)$ is depicted via a solid black line.
    The top panel shows a comparison between a weak and long sinusoidal drive resonant with the ground-first excited state transition and an optimized, much faster control.
    The bottom panel presents an optimal control leading to the GKP state.
    (c) Excited states occupation numbers $\OccupNumbers = |\braket{\FockStateN(\Pos)}{\WaveFunction(\Pos,\Time)}|^2$ for $\GroundState(\Pos) \rightarrow \FockStateOne(\Pos)$ Fock excitation protocols of (b)top.
    The left panel corresponds to the slow Rabi flop, while the right one shows the fast optimal control.
    Utilizing more states within the nonharmonic potential accelerates and makes the control more versatile.
    }
    \label{fig1}
\end{figure}

The preparation of a continuous-variable system in a non-Gaussian quantum state is of paramount importance in various aspects of quantum science.
This ranges from fundamental tests of quantum mechanics~\cite{Zurek2001,Deleglise2008, Arndt1999, Fein2019, Bild2023}, through the design of quantum sensors~\cite{Peters2001, Graham2013a, Parker2018, Margalit2021}, to quantum information processing~\cite{Gottesman2001, Knill2001,Heeres2017, Reinhold2020, Ma2020, Grimm2020, Gertler2021, deNeeve2022, Gonzalez-Cuadra2023}.
The generation of non-Gaussian states requires a nonlinear resource, often introduced through coupling to an auxiliary degree of freedom, e.g., a two-level system~\cite{Cirac1996,Meekhof1996,Heeres2017,Loredo2019,Grimm2020,Trivedi2020,Bild2023}.
On the other hand, some continuous-variable systems already possess intrinsic nonharmonicity in the potential of a canonical variable (see Fig.~\ref{fig1}).
Notable examples include the mo- \noindent tion of a particle in a trap of finite depth~\cite{Jauregui2001,Home2011} and the current in an electric circuit with a Josephson junction~\cite{Blais2021}.
These nonharmonicities in the potential are typically used to define a qubit within continuous-variable systems~\cite{Gorman2005,Gollub2006,Koch2007,Schreier2008,Pistolesi2021}.
In contrast, we investigate whether this intrinsic nonlinearity suffices to implement quantum protocols beyond the two-dimensional subspace. Since it is weaker than the nonlinearity provided by auxiliary two-level systems, we aim to assess how far optimal control can compensate for this limitation.

More specifically, we develop protocols to generate a plethora of states, including Fock, Gottesman-Kitaev-Preskill (GKP)~\cite{Gottesman2001}, multi-legged-cat~\cite{Zurek2001,Leghtas2013a}, and cubic-phase~\cite{Gottesman2001,Zheng2021}, using optimal control~\cite{Glaser2015,Koch2022,Rossignolo2023} of the potential's position and depth.
Furthermore, we employ this control mechanism to design protocols that implement arbitrary unitaries in selected subspaces, enable single-shot orthogonal state discrimination~\cite{Schuster2007,Mallweger2023}, spatial state transfer, correct for non-Gaussian evolution, and perform algorithmic cooling~\cite{Popp2006,Shaw2025}.
These protocols could be implemented with single atoms in optical tweezers~\cite{Serwane2011,Serwane2011a,Kaufman2014,Kaufman2015a,Brown2023a} and lattices~\cite{Jauregui2001,Lam2021} to extend the motional states available in protocols with itinerant particles, e.g., fermionic quantum processors \cite{Gonzalez-Cuadra2023}; or with flux-tunable transmons~\cite{Koch2007,Hutchings2017,Blais2021} for minimally invasive state manipulation [cf. Fig.~\ref{fig1}(a)].
Previous works, including control of Bose-Einstein condensates~\cite{Hohenester2007,Bucker2011,Bucker2013a,Jager2014,vanFrank2016,Hocker2016,Sorensen2018,Schmidt2020,Dupont2021,Xu2022b}, fast atom transport \cite{Calarco2004,Dorner2005,Lam2021}, ion shuttling \cite{Sterk2022},
cat state creation through two-photon driving \cite{Puri2017}, and lattice interferometry \cite{Weidner2018} have demonstrated similar concepts.
However, the role of nonharmonic potentials in quantum information processing remains underexplored, despite their natural relevance---particularly for exploiting, for example, the motion of single atoms in scalable optical tweezer arrays~\cite{Shaw2025}.

The preparation of continuous-variable modes in quantum non-Gaussian states and their subsequent manipulation have been an object of increasing interest across several platforms, including single neutral atoms~\cite{Brown2023a}, trapped ions~\cite{Fluhmann2019,McCormick2019,Podhora2022,Matsos2024}, superconducting circuits~\cite{Wang2019,Eickbusch2022,Kudra2022}, propagating light at the telecommunication wavelength~\cite{Endo2023,Konno2024}, etc.
The control scheme we present offers a realization of a large variety of relevant tasks and can be applied to a generic continuous-variable platform that involves nonharmonicity.
Our protocols take advantage of the controllability of the external potential landscape, naturally offered by, e.g., optical~\cite{Lam2021,Brown2023a,Shaw2025}, electric~\cite{Home2011}, and magnetic~\cite{GutierrezLatorre2023,Hofer2023,Fuwa2023} potentials, and their hybrid combinations~\cite{Fluhmann2019,McCormick2019,Podhora2022,Matsos2024,Bonvin2024a}.
With such schemes, a high information capacity of bosonic degree of freedom is activated via preparation and control of high-energy and high-quality quantum non-Gaussian states.
Our protocols are particularly interesting and useful for systems without well-controlled access to nonlinear resources, e.g., internal degrees of freedom or auxiliary superconducting circuits.
However, they can also be combined with such couplings for simultaneous control of both bosonic and spin degrees of freedom, opening new possibilities for, e.g., the use of hyperentanglement~\cite{Deng2017}.

The paper is structured as follows.
Section~\ref{sec:model} introduces details about the physical model and optimization techniques, while Section~\ref{sec:protocols} presents several protocols that can be achieved through potential modulation, including state preparation, correcting for non-Gaussian evolution, implementation of unitaries, single-shot measurements, cooling, and spatial state transfer.
Further, Section~\ref{sec:feasibility} analyzes the feasibility of our proposal, addressing the speed limits of the protocols, their robustness against decoherence, and how couplings to other modes can be incorporated into optimization.
Section~\ref{sec:conclusions} concludes the paper with a finishing discussion and an outlook.

\section{Control of potential landscape}\label{sec:model}
In the following Section, we present two exemplary realizations of one-dimensional continuous-variable systems for which our proposal can be applied---a single neutral atom trapped via optical forces, either in a tweezer or in a lattice, and a flux-tunable transmon.
In the Subsection~\ref{diffhams}, we provide details of the involved Hamiltonians in these platforms, introduce a universal description of our control scheme, and address how much nonharmonicity is needed for the presented protocols.
The Subsection~\ref{control_optimization} is dedicated to details on the control optimization of our scheme.

\subsection{Hamiltonians for different systems}\label{diffhams}
As a starting point, we focus on a one-dimensional continuous-variable system described by two conjugate quadrature operators, which may correspond to an arbitrary realization of a single mode, e.g., the motion of a single atom in an optical tweezer~\cite{Brown2019,Kaufman2021,Brown2023a} or a lattice~\cite{Robens2018,Lam2021}, or phase and charge operators in a flux-tunable transmon~\cite{Rol2019,Rol2020,Blais2021}.
In the former case, coherent dynamics is driven by a Hamiltonian,
\begin{align}\label{hamDim1}
\Hamiltonian = \frac{\MomentumOperator^2}{2 \Mass} + \left[1 + \InControl(\DimTime)\right] \Pot \left[\PositionOperator -  \ControlPos (\DimTime)\right],
\end{align}
where $\Mass$ is the mass of an atom, $\PositionOperator$ is the position operator, $\MomentumOperator$ is the momentum operator, $\DimTime$ is time, and functions $\InControl(\DimTime)$ and $\ControlPos (\DimTime)$ control the depth and position of the potential.
Here, the trapping potential $\Pot(\PositionOperator)$ is Gaussian for the tweezer,
\begin{align}
\Pot(\PositionOperator) = \TweezerDepth\left(1 - e^{-\frac{2 \PositionOperator^2}{\TweezerWaist^2}}\right),
\end{align}
and a squared sine for the lattice,
\begin{align}
\Pot(\PositionOperator) = \TweezerDepth \sin^2 \left( \frac{2 \pi}{\WaveLength}\PositionOperator\right),
\end{align}
where $\TweezerWaist$ is the tweezer waist, $\WaveLength$ is the optical wavelength, and $\TweezerDepth$ is either the tweezer or the lattice depth.
For the transmon case, the Hamiltonian reads
\begin{align}\label{hamDim2}
\Hamiltonian = 4 \ChargeEnergy \ChargeNumberOperator^2 + & \JosEnergyTot \cos \left( \frac{\pi \ExternalFlux (\DimTime)}{\FluxQuantum} \right)  \sqrt{1+\JunctionAssymetry^2 \tan^2 \frac{\pi \ExternalFlux (\DimTime)}{\FluxQuantum}} \nonumber \\ \times & \left [1-\cos\left(\PhaseDifferenceOperator - \JunctionAssymetry \tan \frac{\pi \ExternalFlux (\DimTime)}{\FluxQuantum} \right)\right].
\end{align}
Here, the phase difference operator is
\begin{equation}
\PhaseDifferenceOperator = \frac{\PhaseDifferenceOperator_1 + \PhaseDifferenceOperator_2}{2},
\end{equation}
where $\PhaseDifferenceOperator_i = \frac{2\pi \PhaseDifferenceOperatorUn_i}{\FluxQuantum}$ denotes the phase difference across the $i^{\text{th}}$ Josephson junction, and $\FluxQuantum = h / 2e$ is the magnetic flux quantum.
The charge number operator is defined as
\begin{equation}
\ChargeNumberOperator = \frac{\ChargeOperator}{2e},
\end{equation}
and the charge energy is
\begin{equation}
\frac{\ChargeEnergy}{h} = \frac{e^2}{2h \Capacitance},
\end{equation}
where $\Capacitance$ is the total capacitance.
The total Josephson energy is
\begin{equation}
\JosEnergyTot = \JosOne + \JosTwo,
\end{equation}
with $\Josi$ representing the Josephson energy of the $i^{\text{th}}$ junction. The asymmetry between the two junctions is characterized by
\begin{equation}
\JunctionAssymetry = \frac{\JosTwo - \JosOne}{\JosTwo + \JosOne}.
\end{equation}
Finally, $\ExternalFlux(\DimTime)$ denotes the time-dependent external magnetic flux threading the SQUID loop.

In the leading order, each of these setups is harmonic with frequency $\Frequency$, yielding natural time $\Time = \Frequency \DimTime$, canonical length and momentum scales,
\begin{align}
\PosOp & = \frac{\PositionOperator}{\OscillatorLength} = \frac{\ChargeNumberOperator}{\ChargeLength} =\frac{1}{\sqrt{2}} (\Creation + \Annihilation), \nonumber \\
\MomOp &= \frac{\MomentumOperator}{\MomentumLength} = \frac{\PhaseDifferenceOperator}{\PhaseLength} =\frac{1}{\sqrt{2}}  i(\Creation - \Annihilation),
\end{align}
where $\Creation$ creates a single excitation and $[\PosOp,\MomOp]=i$.
For an atom trapped in a tweezer, we have 
\begin{align}
\OscillatorLength &=(\hbar^2 \TweezerWaist^2 / 4 \Mass  \TweezerDepth )^{1/4}, \nonumber \\
 \MomentumLength &= \hbar / \OscillatorLength, \nonumber \\
 \Frequency &= \sqrt{4 \TweezerDepth / \Mass \TweezerWaist^2},
\end{align}
while for the optical lattice, 
\begin{align}
\OscillatorLength &=(\hbar^2 \WaveLength^2 / 8 \pi^2 \Mass  \TweezerDepth )^{1/4}, \nonumber \\
 \MomentumLength &= \hbar / \OscillatorLength, \nonumber \\
 \Frequency &= \sqrt{8 \pi^2 \TweezerDepth / \Mass \WaveLength^2}.
\end{align}
For the flux-tunable transmon, the canonical length and momentum scales read 
\begin{align}
\ChargeLength &=( 8 \ChargeEnergy / \JosEnergyTot)^{1/4}, \nonumber \\
\PhaseLength &= 1 / \ChargeLength, \nonumber \\
\Frequency &= \sqrt{8 \ChargeEnergy \JosEnergyTot } / \hbar.
\end{align}

In the canonical variables $\PosOp$, $\MomOp$, each of the Hamiltonians~\eqref{hamDim1},~\eqref{hamDim2} can be rewritten as
\begin{equation} \label{Eq:1}
   \frac{\Hamilton} {\hbar \Frequency} = \frac{1}{2} \MomOp^2  + \left[1 + \InControl(\Time)\right] \DimPotNon  \left[   \PosOp - \PosControl(\Time) \right] , 
\end{equation}
where $\DimPotNon = \Pot / \hbar \Frequency$ is the dimensionless potential.
The control of the system is performed through optimal modulation of position, $\PosControl(\Time)$, and depth, $\InControl(\Time)$, of the potential.
While for an atom in an optical tweezer, they are independent controls realized through, e.g., acousto-optic modulator~\cite{Kaufman2021}, for flux-tunable transmons, they are constrained through a single control function, the intensity of the external flux~\cite{Rol2019},
\begin{align}\label{flux_con}
\InControl(\Time) &= \sqrt{\frac{1+4 \sqrt{\ChargeEnergy/2 \JosEnergyTot} \PosControl^2(\Time)}{1+4 \sqrt{\ChargeEnergy/2 \JosEnergyTot} \PosControl^2(\Time) \JunctionAssymetry^{-2}}}-1, \nonumber \\
\PosControl(\Time) &= \frac{\JunctionAssymetry}{2 \NonLin} \tan \frac{\pi \ExternalFlux (\Time)}{\FluxQuantum}.
\end{align}
Such modulations have been realized in various other experimental platforms~\cite{Bucker2013,Weidner2018,Lam2021,Schmidt2020,Dupont2021}, where the choice of either position or depth control depends on feasibility within a specific setup.
For example, Paul traps allow for both position and depth control via varying current through electrodes generating the electric field~\cite{Schmidt2020}, while in optical lattices, phase control allows very precise position control~\cite{Lam2021}.
As for optical tweezers, the use of multiple radiofrequency tones makes a powerful knob for the time-dependent control of both position and depth~\cite{Brown2023a}.

In general, we consider the shape of the potential to be symmetric in $\Pos$, and we find it useful to express it as 
\begin{align}
\DimPotNon(\Pos) = \frac{1}{2 \NonLin^2}\DimPot (\NonLin \Pos)  \approx  \frac{1}{2}\Pos^2 -  \frac{1}{6} \NonLin^2  \Pos^4
\end{align}
when expanded up to the second order in the small parameter $\NonLin$.
Note that, as the potential $\DimPotNon(\Pos)$ is beyond quadratic in $\Pos$, the ensuing dynamics is \textit{nonlinear}, i.e., it does not preserve the Gaussianity of the states.
For the examples considered in this work, the tweezer potential then reads as
\begin{align}
\DimPot (\Pos) = \frac{3}{2} \left[1 - \exp\left(-\frac{2}{3}\Pos^2\right)\right],
\end{align}
while for the lattice it is 
\begin{align}
\DimPot (\Pos)  = \sin^2 \Pos,
\end{align}
and for the flux-tunable transmon,
\begin{align}
\DimPot (\Pos) =  \frac{1}{2} \left[1-\cos \left( 2 \Pos \right) \right]
\end{align}
Note that we have defined the units so that the functional forms of these potentials match up to the second leading order, allowing for a unified discussion of nonharmonicity.
Here, $\NonLin$ is the measure of the nonharmonicity of the potential, comparing the canonical length to the characteristic length scale of the potential, and translates directly into an energy level splitting, 
\begin{align}
\frac{\Frequency_{21} - \Frequency_{10}}{\Frequency} = -\frac{1}{2}\NonLin^2,
\end{align}
where $\Frequency_{ij}$ is a transition frequency between $i^{\text{th}}$ and $j^{\text{th}}$ levels.
For instance, for an atom in a tweezer, it is 
\begin{align}
\NonLin = \sqrt{3} \frac{\OscillatorLength}{\TweezerWaist};
\end{align}
for an optical lattice, it is a Lamb-Dicke parameter
\begin{align}
\NonLin = \frac{2 \pi \OscillatorLength}{\WaveLength};
\end{align} for the flux-tunable transmon, it is a monotonic function of transmon anharmonicity $\AnHarm$, 
\begin{align}
\frac{\AnHarm}{ \Frequency} = \sqrt{\frac{\ChargeEnergy}{8 \JosEnergyTot}} =  \frac{1}{2}\NonLin^2;
\end{align}
and for Kerr oscillators, it can be associated with Kerr nonlinearity $\Kerr \sim \Frequency \NonLin^2 /24$~\cite{Dykman2012}.
The typical values of physical parameters for the examples introduced in this section lead to nonharmonicities no larger than $\NonLin \sim 0.5$ (see Tab.~\ref{tab1}) and can be tuned down at least an order of magnitude depending on the specific setup.
Hence, here we will analyze such a parameter regime.
Nevertheless, many other experimental platforms are characterized by much lower values, such as trapped ions~\cite{Leibfried2003,Home2011} or levitated mechanical oscillators~\cite{Gonzalez-Ballestero2021,Roda-Llordes2024b}.
There, lower nonharmonicity is associated with either larger potential length scales or larger masses of trapped objects. 
In such cases, the additional enhancement of nonharmonicity is needed, e.g., through state excitation~\cite{Lingenfelter2021,Yuan2025,Roda-Llordes2024b}, to generate non-Gaussian states.

\begin{table}[h!]
\begin{tabular}{m{4.3em} m{5em} |m{4.3em} m{5em} }
\hline\hline
\multicolumn{2}{c|}{Optically trapped atom}         & \multicolumn{2}{c}{Superconducting circuit}        \\ \hline
$\Mass$                          & $10^1$-$10^2$ u  & $\ChargeEnergy/ h$   & MHz-GHz                     \\
$ \TweezerWaist$, $ \WaveLength$ & $10^2$-$10^3$ nm & $\JosEnergyTot$      & $ 10^1$-$10^4\ChargeEnergy$ \\
$\TweezerDepth/\Boltzmann$       & $\mu\text{K}$-mK & $\JunctionAssymetry$ & 0-1                         \\ \hline\hline
\end{tabular}
\caption{Typical physical parameters for atoms held in optical tweezers or optical lattices and flux-tunable transmons.\label{tab1}}
\end{table}

\subsection{Control optimization}\label{control_optimization}

Before presenting specific protocols, let us discuss possible methods of designing controls $\PosControl(\Time)$ and $\InControl(\Time)$.
Within each of the subsequent protocols, we aim to achieve some specific goal---be it state preparation, unitary implementation, or others---that can be quantified through a reward function that ought to be maximized.
A set of methods for designing time-varying controls for the maximization of such a reward function is collectively called \textit{quantum optimal control} (QOC)~\cite{Werschnik2007,Glaser2015,Koch2022}.
Introduced in the eighties and since then broadly developed, QOC has become one of the main quantum control tools, among other optimization approaches, such as adiabatic passages~\cite{Vitanov2017}, shortcuts to adiabaticity~\cite{Guery-Odelin2019}, and composite pulses~\cite{Levitt1986}. 
Various techniques have been established to design the control pulses, relying on both gradient-free and gradient-based approaches.
The latter assume the ability to differentiate the reward function and include, among others, GRAPE (gradient ascent pulse engineering)-~\cite{Khaneja2005} and Krotov-like~\cite{Reich2012,Goerz2019} techniques for optimizing controls that are piecewise constant in time.
On the other hand, the former rely only on the evaluation of the reward function itself. 
The examples are numerous, including Nelder-Mead~\cite{Nelder1965}, evolution strategies~\cite{Goldberg2007}, simulated annealing~\cite{Kirkpatrick1983}, and many others.
Notably, recent years have brought a rapid surge in the use of machine learning approaches for quantum control problems~\cite{Giannelli2022}.
The choice of approach relies on constraints given by a specific experimental setup, including how accurately and fast we can solve the dynamics and what the limitations of control pulses are, such as maximal Fourier bandwidth and intensity.
We choose to utilize the dressed chopped random-basis technique (dCRAB)~\cite{Muller2022} that is based on the Nelder-Mead gradient-free method.
Here, the control function is expanded into a Fourier basis with randomized frequencies and a high-frequency cutoff corresponding to approximations of experimentally accessible bandwidths, i.e.,
\begin{align}
    \PosControl(\Time) = \sum_{k=1}^{N_{\text{p}}} a_k \cos \nu_k \Time + b_k \sin \nu_k \Time,
\end{align}
where $a_k$ and $b_k$ are parameters to be optimized, $\nu_k$ are frequencies that are probabilistically drawn from the uniform distribution in the range $[0,\nu_\text{max}]$~\cite{Caneva2011}, $\nu_\text{max}$ is the frequency cutoff, and $N_{\text{p}}$ is the number of frequency components.
We have chosen this particular gradient-free method as we consider several different reward functions, it naturally implements a frequency cutoff for the control function, and there are available well-developed open-source packages, including the Quantum Optimal Control Suite~\cite{Rossignolo2023}, which we utilize for the optimization.
However, depending on the specifics of the experimental implementation, our results can be achieved with an alternative approach, such as, e.g., GRAPE.

Generally speaking, controllability of the system described by Eq.~\eqref{hamDim1} depends sensitively on the shape and depth of the potential.
For shallow potentials, the finite number of bound states can limit the set of achievable state transformations.
However, in the case of sufficiently deep or unbounded potentials---such as the quartic potential---full controllability can be established under certain control schemes~\cite{Zhao2023}.
The chosen potential landscape is then crucial in determining the accessible Hilbert space and, consequently, the effectiveness of optimal control protocols.

For all the simulations, the split-step method~\cite{Leforestier1991} was used to simulate the dynamics on a spatial grid.
In the case of one-dimensional calculations, throughout optimization runs, the numerical parameters read: number of spatial grid points $N_x = 2^8$, where the grid spanned $x \in [-13,13]$, and number of time grid points $N_\Time = 500$.
After optimizing the pulses, the dynamics was run and further optimized with increased accuracy to make sure the results are converged.
The optimization has been performed with QuoCS library version 0.0.44~\cite{Rossignolo2023}.
The number of Fourier components used for dCRAB varied from 20 to 50, depending on a particular protocol, and was randomized from a uniform distribution on a bandwidth that also varied between different protocols.
The choice of these optimization parameters was fine-tuned for each of the protocols, however, it was not exclusive---different choices usually yielded similar values of specific reward functions, but with different total optimization iterations.
Note that further optimization of all the protocols towards lower infidelities is possible and would involve further fine-tuning of optimization parameters.

\section{Protocols}\label{sec:protocols}
In the following section, we discuss several protocols that can be implemented via a nonharmonic potential modulation.
It includes a versatile Gaussian (Sec.~\ref{prot:gaussian}) and non-Gaussian state preparation, both in a single- (Sec.~\ref{prot:state_sw}) and double-well (Sec.~\ref{prot:state_dw}) potential landscape, as well as the implementation of unitaries (Sec.~\ref{prot:unitaries}), single-shot measurements (Sec.~\ref{prot:state_meas}), cooling (Sec.~\ref{prot:cooling}), and correcting state evolution for nonlinear evolution due to nonharmonicities (Sec.~\ref{prot:stab}).

\begin{figure}[h!t!]
    \includegraphics[width=0.97\linewidth]{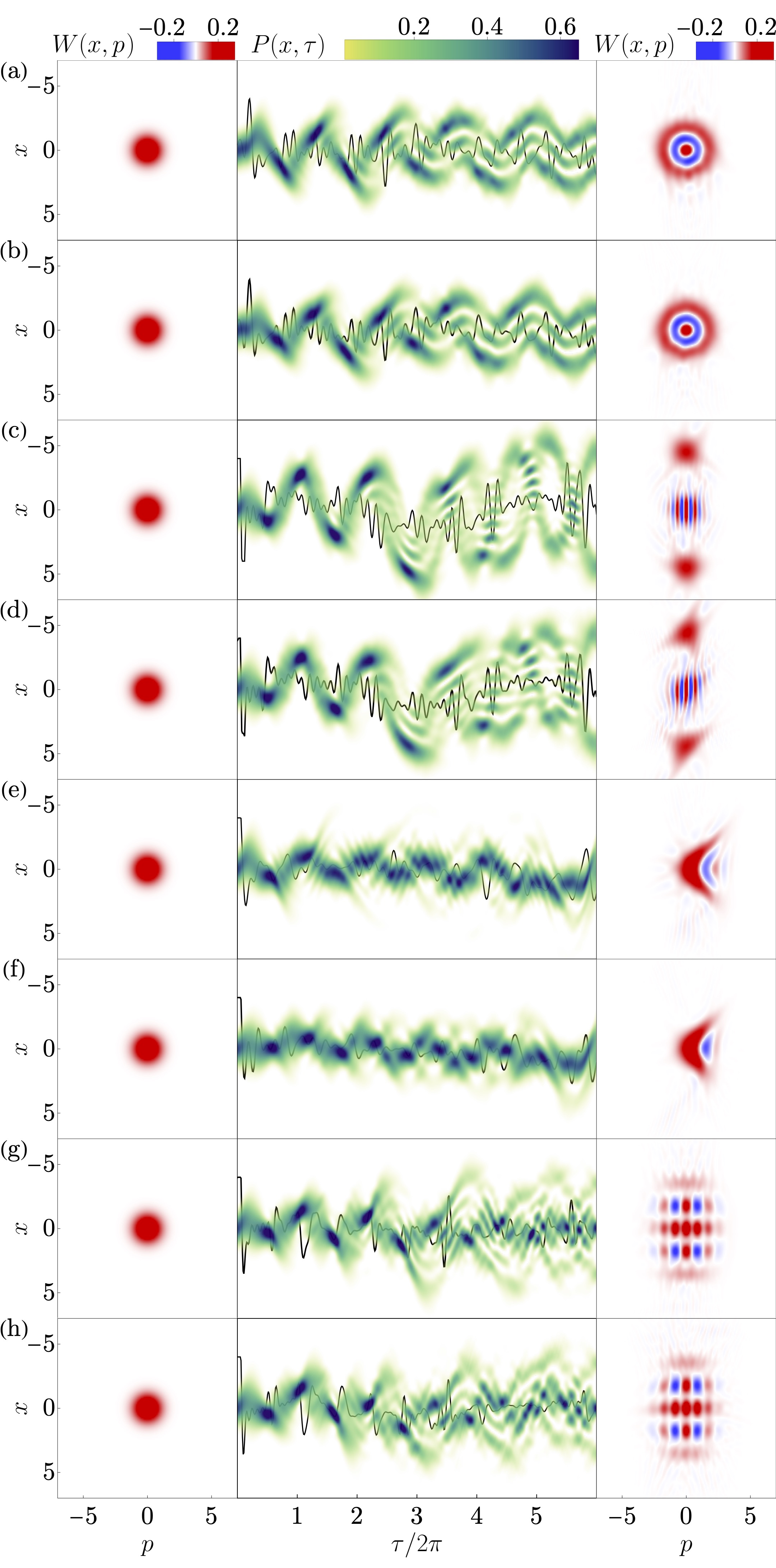}
    \caption{Examples of a state preparation protocol involving Gaussian [(a), (c), (e), (g)] and cosine [(b), (d), (f), (h)] single-well potentials, with $\NonLin = 0.25$ and $\TimeMax/ 2 \pi = 6$, using only optimized position of the potential $\PosControl(\Time)$ in the case of the Gaussian potential.
    The protocols include second excited Fock [(a-b)], cat [(c-d), $\TweezerSeparation = 9$], cubic-phase [(e-f), $ \Cubicity = 2$, $\Squeeze = 0.7$], and GKP [(g-h), $\Squeeze = 0.6$, $(\Disp_1,\Disp_2,\Disp_3) = (-\sqrt{4 \pi},0,\sqrt{4 \pi})$, $(\coeff_1,\coeff_2,\coeff_3) = (1,2,1)$] state preparation.
    Protocol fidelities $\Fidelity$  are: (a) $99.8 \%$ (b) $99.8 \%$ (c) $92.7 \%$ (d) $93.4 \%$ (e) $97.2 \%$ (f) $94.9 \%$ (g) $99.8 \%$ (h) $99.8 \%$. See the caption of Fig.~\ref{fig1} for subplot and curve legend details.
    \label{fig2}}
\end{figure}

\subsection{State preparation}\label{prot:state_sw}
As the first type of protocol, we consider state preparation.
The initial state of the system is assumed to be the ground state $\GroundState(\Pos)$ of the nonharmonic potential $\DimPotNon(\Pos)$, well approximated by the wave function
\begin{align}
    \GroundState(\Pos) \approx \frac{1}{ \pi^{1/4}} e^{- {\Pos^2/2}},
\end{align}
which is the ground state of the leading harmonic approximation, $\Pos^2/2$.
For this protocol, we choose to control only the position of the potential $\PosControl(\Time)$, to maximize the fidelity
\begin{align}
\Fidelity = \left|\braket{\WaveFunction(\Pos,\Time = \TimeMax)}{\TargetState(\Pos)} \right|^2
\end{align}
with the target state $\TargetState(\Pos)$.
We consider the preparation of several states, including Fock, finite GKP, cubic-phase, and cat states with high fidelity in a one-dimensional geometry.
As these states are non-Gaussian, the role of nonharmonicity of the potential is evident---if it was quadratic, and dynamics linear, the Gaussianity of the state would be preserved during the evolution.
In Fig.~\ref{fig1}(b), we present an example of the excitation protocol, where we specialize to a Gaussian potential with $\NonLin = 0.25$.
The target state is a finite GKP state, 
\begin{align}\label{gkp_state}
\GKP(\Pos) = \Normalization \sum_{i=1}^3 \coeff_i e^{\Squeeze/2}  \GroundState [ e^{\Squeeze} ( x - \Disp_i)]
\end{align}
with $ (\coeff_1,\coeff_2,\coeff_3) = (1,2,1)$, $(\Disp_1,\Disp_2,\Disp_3) = (-\sqrt{4 \pi},0,\sqrt{4 \pi})$, the squeezing parameter $\Squeeze = 0.6$, the normalization constant $\Normalization$, and the whole protocol is set to take $\TimeMax /2 \pi = 6$.
The presented protocol yields fidelity $\Fidelity \approx 99.8\%$.
The further examples of the state preparation are presented in Fig.~\ref{fig2}, both for the Gaussian potential and for cosine potential with the constraint~\eqref{flux_con}, $\NonLin = 0.25$ and $\JunctionAssymetry = 0.8$.
Specifically, we additionally show the second excited Fock state,
\begin{align}
    \WaveFunction_2(\Pos) = \frac{1}{\sqrt{2}} \frac{1}{\pi^{1/4}} \left(2 x^2-1\right) e^{-x^2/2},
\end{align}
the squeezed cubic-phase state,
\begin{align}
    \WaveFunction_\text{cub}(\Pos) = e^{i \Cubicity x^3} e^{\Squeeze/2} \GroundState\left( e^{\Squeeze} x \right), 
\end{align}
and the cat state,
\begin{align}
    \WaveFunction_\text{cat}(\Pos) = \frac{1}{\sqrt{2}} [ \GroundState(\Pos + \TweezerSeparation/2) + \GroundState(\Pos - \TweezerSeparation/2) ], 
\end{align}
where $\Cubicity$ is the cubicity, $\Squeeze$ is the squeezing parameter, and $\TweezerSeparation$ is the separation between the coherent states.
The specific values of the state parameters and fidelities are shown in the caption.

\subsection{Gaussian operations}\label{prot:gaussian}
\begin{figure}[h!t!]
    \includegraphics[width=\linewidth]{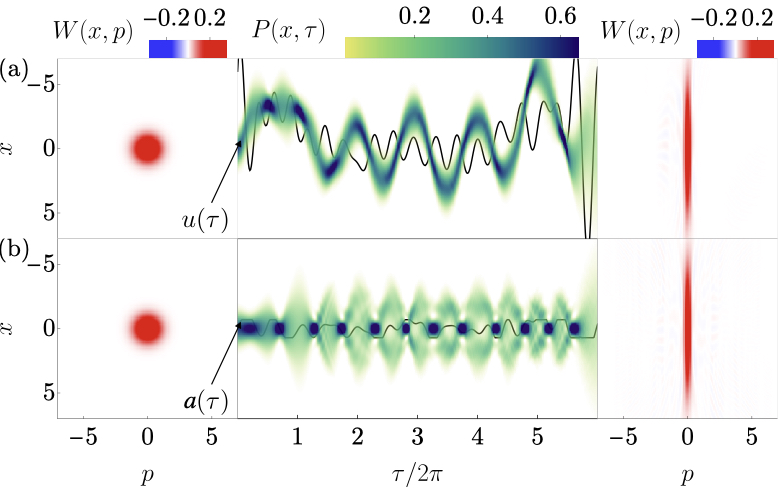}
    \caption{Gaussian squeezing of the initial ground state via (a) displacement and (b) modulation. The potential is taken to be Gaussian with $\NonLin=0.15$. Protocols fidelities read: (a) $99.8 \%$ (b) $99.6 \%$. See the caption of Fig.~\ref{fig1} for subplot and curve legend details.
\label{fig3}}
\end{figure}

As we consider nonlinear dynamics, a relevant question is whether it is still possible to implement Gaussian operations within this framework.
In our setting, this typically involves protocols that transiently generate non-Gaussian states during the evolution, but yield a final state that is Gaussian.
A relevant example is the preparation of squeezed vacuum states, for which we demonstrate two distinct realizations.
The first protocol employs only displacement modulation, while the second relies solely on depth modulation.
Both cases are shown in Fig.~\ref{fig3} and achieve high final fidelities.
The second example represents a generalized version of the frequency-jump protocol, in which a sudden change in potential depth excites breathing in the wave function, resulting in squeezing.
In our case [Fig.~\ref{fig3}(b)], the achieved squeezing surpasses what would be expected from a single-frequency-jump protocol, which is fundamentally limited by the ratio of initial and final frequencies.

\begin{figure}[ht!]
    \includegraphics[width=\linewidth]{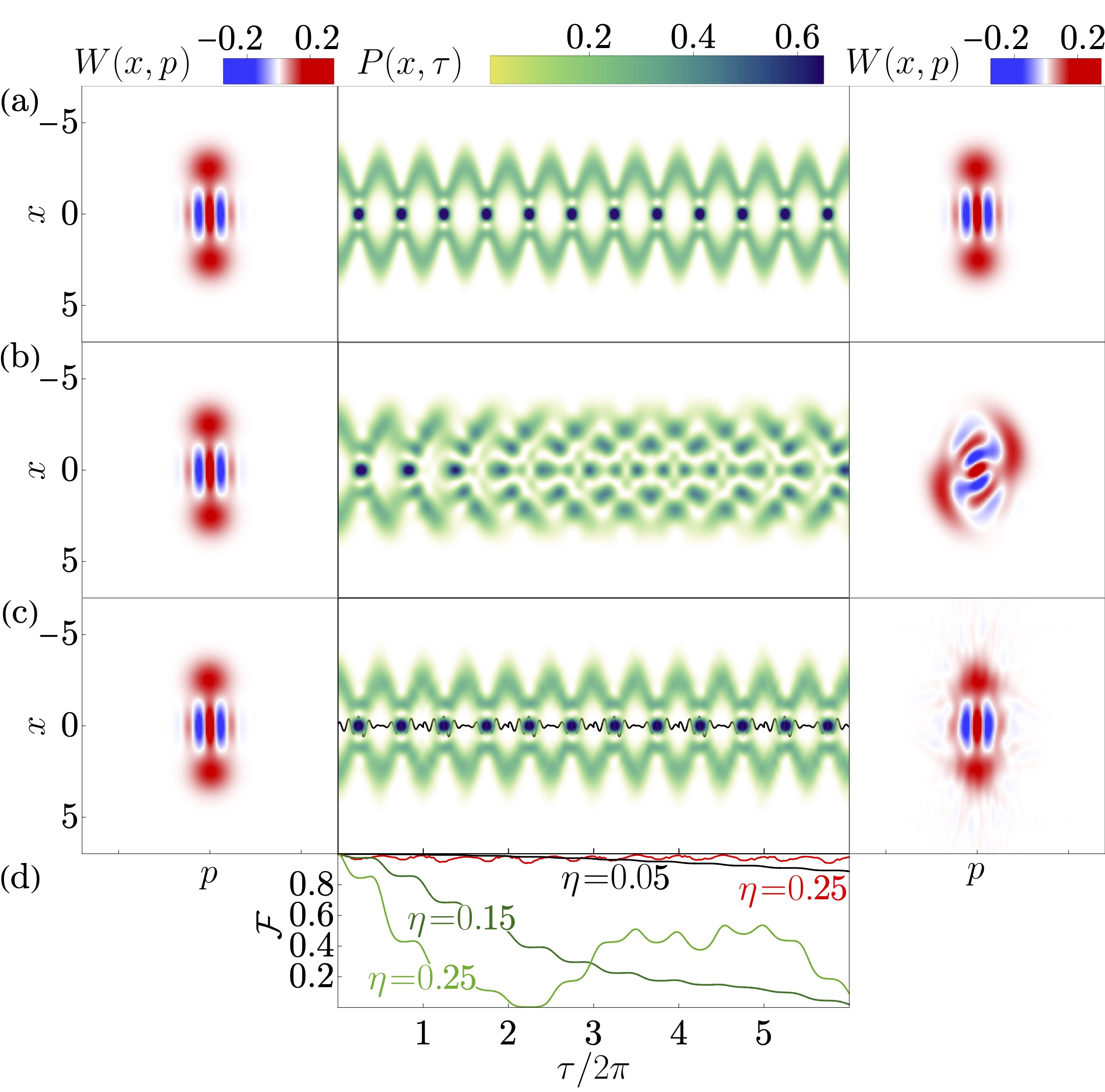}
    \caption{ (a) Temporal evolution of a cat state in a perfectly harmonic potential.
    It rotates in the phase space and fully revives every trap period.
    (b) Evolution of a cat state in a tweezer characterized by $\NonLin = 0.25$.
    (c) Evolution of a cat state in a tweezer characterized by $\NonLin = 0.25$, however, with optimally controlled potential position and depth.
    (d) Fidelities between a cat state rotating in a perfectly harmonic potential and in Gaussian potentials with different nonharmonicities without (shades of green) and with (red) optimal control. See the caption of Fig.~\ref{fig1} for subplot and curve legend details.  
    \label{fig4}}
\end{figure}

\subsection{State stabilization}\label{prot:stab}
Note that undriven evolution is nonlinear and, hence, the prepared state gets distorted during the evolution. 
In contrast, when evolved in the perfectly harmonic potential, the state undergoes rotation in the phase space with the trap period.
In a realistic setup, such a state preservation in the rotating frame is destroyed by the nonharmonicity of the potential.
The remedy lies in either decreasing the nonharmonicity or, alternatively, mimicking perfect phase space rotation through optimal control~\cite{Puri2017}.
We present the latter method, which consists of two optimization steps.
The first one involves using the average fidelity with respect to the perfectly rotating state during a multiple of the trap period $\TimeMax = n 2 \pi$ as a reward function,

\begin{equation}
\Fidelity_\text{rot} = \frac{1}{\TimeMax}\int \dd \Time  \left|\braket{\WaveFunction(\Pos,\Time)}{\WaveFunction_\text{rot}(\Pos, \Time)} \right|^2, 
\end{equation}
where
\begin{align}
\ket{\WaveFunction_\text{rot} (\Time)} = e^{-i \hat{a}^{\dagger} \hat{a} \Time} \ket{\WaveFunction (0)}
\end{align}
is the initial state undergoing a phase-space rotation in a perfectly harmonic trap.
After optimizing $\Fidelity_\text{rot}$, the next step involves maximizing the final fidelity with the initial state after a single trap period,
\begin{equation}
\Fidelity = \left|\braket{\WaveFunction(\Pos, \TimeMax)}{\WaveFunction(\Pos,0)} \right|^2.
\end{equation}
The second step guarantees high fidelity, should the protocol be applied consecutively many times, while the first one ensures that the state is not distorted at any time.
In Fig.~\ref{fig4}, we present an example of an optimally controlled cat state stabilization.
The optimization is conducted for a single trap period, $\TimeMax = 2 \pi$, with both position and depth control of a Gaussian potential.
Subsequently, the optimized control is applied six times consecutively over six trap periods and compared with the bare evolution in a Gaussian potential.
The results showcase that the presented optimal stabilization of the state surpasses that of a Gaussian potential with a relatively low nonharmonicity $\NonLin = 0.05$, within this particular time scale.

\begin{figure}[h!t!]
    \centering
    \includegraphics[width=\linewidth]{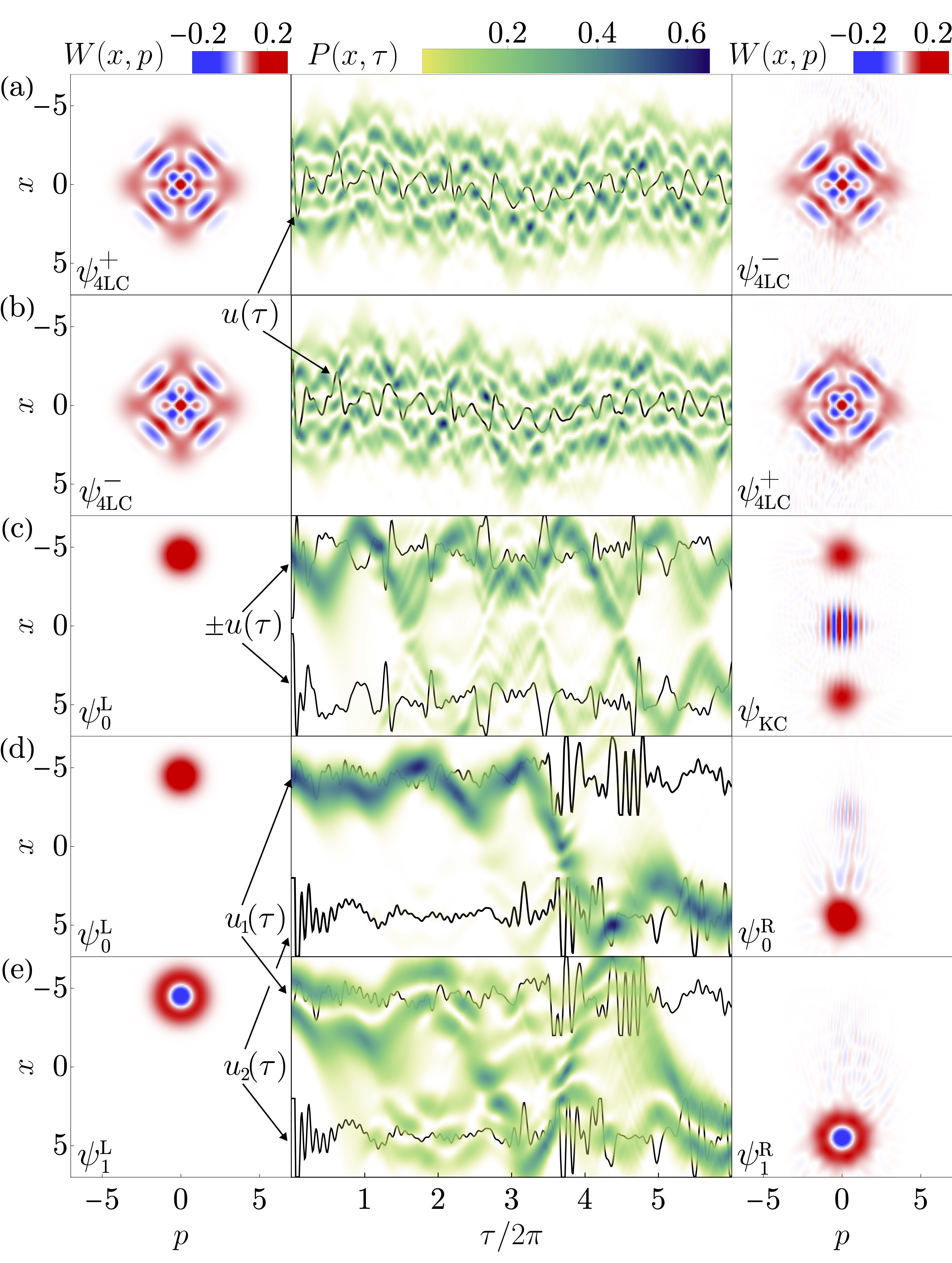}
    \caption{(a,b) Two orthogonal four-legged-cat states spanning a two-level subspace evolved in a position- and depth-controlled single Gaussian well, realizing a $\SigmaX$ operation.
    (c) Cat state preparation using a double-well optimal control.
    The superscript L(R) signifies the state centered in the left(right) potential well.
    (d,e) State transfer between the wells---each of the vibrational states, $\GroundState(\Pos)$ and $\FockStateOne(\Pos)$, is transferred to the other well, without altering the relative phase. See the caption of Fig.~\ref{fig1} for subplot and curve legend details.
   }
    \label{fig5}
\end{figure}

\subsection{Unitaries}\label{prot:unitaries}
\begin{figure}[h!t!]
    \includegraphics[width=\linewidth]{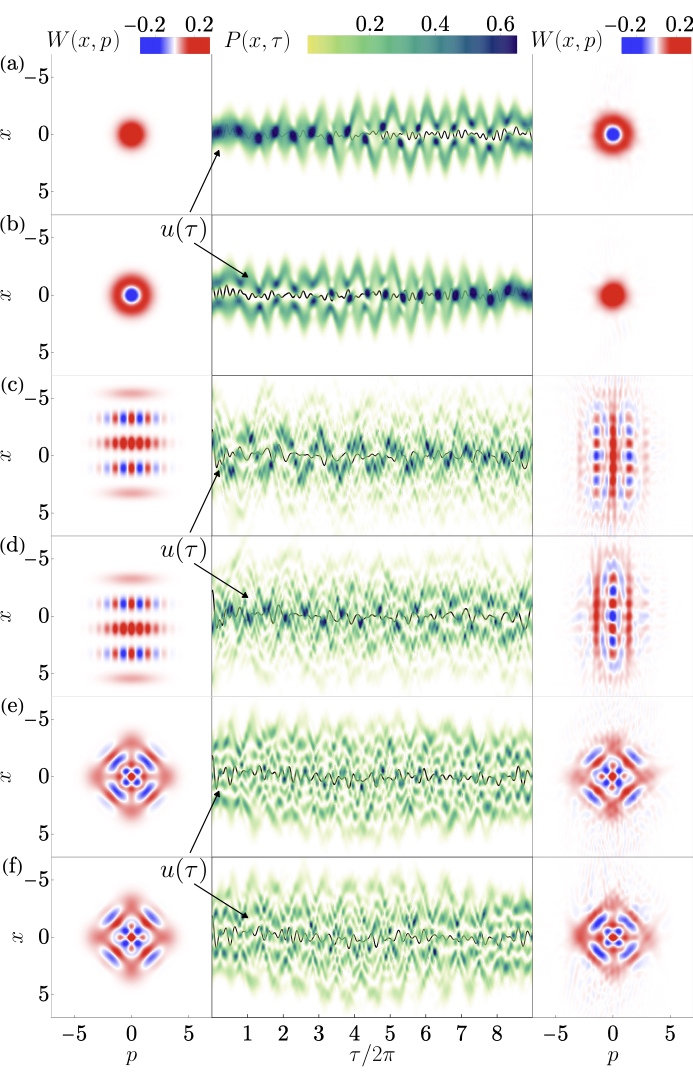}
    \caption{Evolution of selected orthogonal states in a Gaussian potential characterized by $\NonLin = 0.25$ and with optimally controlled displacement $\PosControl(\Time)$ and intensity $\InControl(\Time)$.
    The solid line shows $\PosControl(\Time)$ and the protocol lasts $\TimeMax/ 2 \pi = 9$.
    (a-b) $\SigmaX$ unitary within a subspace spanned by $\GroundState$ and $\WaveFunction_1$ Fock states.
    The fidelity reads $\FidelityGate \approx 99.8 \%$.
    (c-d) Hadamard unitary within a subspace spanned by GKP states with $\Squeeze = 0.7$, $(\Disp_1,\Disp_2,\Disp_3) = (-\sqrt{6 \pi},0,\sqrt{6 \pi})$.
    The fidelity reads $\FidelityGate \approx 96.4 \%$.
    (e-f) $\SigmaY$ unitary within a subspace spanned by four-legged-cat states states with $\CohDis = 2$.
    The fidelity reads $\FidelityGate \approx 96.2 \%$. See the caption of Fig.~\ref{fig1} for subplot and curve legend details.\label{fig6}}
\end{figure}

The next type of protocol involves the implementation of a specific unitary within a selected subspace.
For simplicity and relevance to several proposals and realizations of bosonic qubits~\cite{Ma2021,Rosenblum2018,Reinhold2020,Grimm2020,Ma2020,Gertler2021,Fluhmann2019,deNeeve2022}, we analyze an example of a two-level subspace.
This subspace can be spanned by any two orthogonal states $\WaveFunction^{\pm}(\Pos)$, including two lowest-lying vibrational states $\GroundState(\Pos)$ and $\FockStateOne(\Pos)$ (Fock basis), two mutually displaced GKP states (GKP basis), four-, and two-legged-cat bases.
For the protocol considered, the reward function is the subspace average gate fidelity~\cite{Pedersen2007}
\begin{align}\label{gatefid}
\FidelityGate = \frac{1}{6}[ \Trac \ ( \AuxM \AuxM^{\dagger} ) + |  \Trac \ \AuxM |^2 ],
\end{align}
where 
\begin{align}
\AuxM = \Projector \TargetUnitary \Unitary \Projector,
\end{align}
$\Projector$ is a projector onto a subspace, $\TargetUnitary$ is a target unitary, and $\Unitary$ is a unitary generated through Eq.~\eqref{Eq:1}.
Taking advantage of the optimized position of the potential $\PosControl(\Time)$ and additional slight, optimized modulation of the depth $\InControl(\Time)$ (to increase the fidelity of the protocol), we present examples of unitaries, e.g., $\SigmaX$ or Hadamard operations, for the above-mentioned subspaces.
In Fig.~\ref{fig5}(a,b), we show an exemplary case of four-legged-cat basis and $\SigmaX$ operation performed with a Gaussian potential, where the basis states are given by
\begin{align}
\FourLeggedCat^{\pm} (\Pos) = \NormFourCat_{\pm} \left[ \Coh_{\CohDis} (\Pos) + \Coh_{-\CohDis} (\Pos) \pm  \Coh_{i\CohDis}  (\Pos) \pm  \Coh_{-i\CohDis}  (\Pos)  \right],
\end{align}
respectively~\cite{Leghtas2013,Mirrahimi2014}.
Here, the coherent state reads 
\begin{align}
\Coh_{\CohDis} (\Pos) = \frac{1}{\pi^{1/4}} \exp [ - (x-\sqrt{2} \Re \CohDis )^2/2 + i \sqrt{2} x \Im \CohDis ],
\end{align}
$\NormFourCat_{\pm}$ are the normalization constants, and we use $\CohDis = 2$.
Again, the protocol is performed with high fidelity, $\FidelityGate \approx 99\%$.
In Fig.~\ref{fig6}, we show additional examples for the Gaussian potential, including, among others, the GKP basis, where the basis is spanned by two mutually displaced GKP states,
\begin{align}
\WaveFunction^{\pm} = \GKP(\Pos \pm \Disp_3/4),
\end{align}
where $\Disp_3$ is defined as in~\eqref{gkp_state} and the squeezing and displacement need to be enough for the states not to overlap.
Further unitary implementations, but for the cosine potential with the constraint~\eqref{flux_con}, are presented in the Appendix~\ref{app:gates1}.

\subsection{Double-well potential landscape}\label{prot:state_dw}
Up to now, we have analyzed quantum dynamics taking place in a single well of a potential landscape.
Here, we bring our attention to the nonharmonic potential landscape case that involves two independently controlled potential wells.
Such a realization is available for optical lattices, optical tweezers~\cite{Kaufman2014,Murmann2015a,Islam2015,Kaufman2021,Yan2022}, or various circuit quantum electrodynamics setups~\cite{Frattini2024}.
First, we address the state preparation and single-particle unitary implementation with two potential wells.
We assume that we have two independently controlled wells at our disposal, such that the total potential reads
\begin{align}\label{twowells}
\DimPotNon(\Pos) = \sum_{i=\{1,2\}} \frac{1}{ 2 \NonLin^2}\DimPot [ \NonLin ( \Pos - \PosControl_i(\Time) ) ]
\end{align}
with two independent position control functions, $\PosControl_i(\Time)$.
Here, we assume that $\DimPot$ contains only a single well with some characteristic width.
Controlling the relative distance between the wells amounts to changing the barrier height, which affects the coupling between the bound states in each of the wells.
It can be understood as mode multiplexing~\cite{Martinez-Garaot2013}, accelerated through the optimal control.
In Fig.~\ref{fig5}(c), we show a balanced cat state preparation utilizing two Gaussian potential wells,
\begin{align}
\GroundState(\Pos \pm \TweezerSeparation / 2) \rightarrow \frac{1}{\sqrt{2}}[ \GroundState(\Pos \pm \TweezerSeparation / 2) + i \GroundState(\Pos \mp \TweezerSeparation / 2 )],
\end{align}
where $\TweezerSeparation$ is the separation between the wells.
Note that the presented control is symmetric, $\PosControl_1(\Time) = - \PosControl_2(\Time) = \PosControl(\Time)$, and the instantaneous separation between the wells is constrained in the optimization.
The fidelity of the presented protocol yields $\Fidelity \approx 99\%$.
The cat states produced in such a manner are pertinent to the fundamental tests of quantum mechanics, especially for massive objects. 
This protocol, along with a fast optimized transport~\cite{Lam2021}, should allow for a fast macroscopic cat state creation and interferometric protocols.

The double-well potential has also been shown to provide a platform for fault-tolerant quantum computing with so-called Kerr-cats~\cite{Grimm2020}.
There, the computational subspace is spanned by superpositions of the ground states of the respective wells, corresponding to the (almost) degenerate ground-state manifold of a full double-well potential,
\begin{align}
\KerrCat^\pm (\Pos) =\frac{1}{\sqrt{2}} [ \GroundState(\Pos + \TweezerSeparation/2) \pm \GroundState(\Pos - \TweezerSeparation/2) ].
\end{align}
With optimal control, one can again realize arbitrary unitaries within this subspace, utilizing the subspace average gate fidelity~\eqref{gatefid}.
Examples of $\SigmaX$, $\SigmaY$, and Hadamard unitaries are presented in detail in Appendix~\ref{app:gates2}.
Furthermore, independent control of two wells can also be used to perform a state-preserving transport between the wells~\cite{Beugnon2007}.
Here, the initial state is prepared within a two-dimensional subspace spanned by two orthogonal states localized in the left (L) well,
\begin{align}\label{leftwell}
    \WaveFunction^\text{L}_\pm (\Pos)= \WaveFunction^\pm (\Pos + \TweezerSeparation/2).
\end{align}
The aim of the protocol is to transfer such a state to a subspace localized in the right (R) well, 
\begin{align}\label{righttwell}
    \WaveFunction^\text{R}_\pm (\Pos)= \WaveFunction^\pm (\Pos - \TweezerSeparation/2).
\end{align}
Then, if we introduce the following matrix,
\begin{align}
\AuxM' = \begin{pmatrix}
\braket{\WaveFunction^\text{L}_{+}(\Pos, \TimeMax)}{\WaveFunction^\text{R}_{+}(\Pos)} & \braket{\WaveFunction^\text{L}_{+}(\Pos, \TimeMax)}{\WaveFunction^\text{R}_{-}(\Pos)} \\
\braket{\WaveFunction^\text{L}_{-}(\Pos, \TimeMax)}{\WaveFunction^\text{R}_{+}(\Pos)} & \braket{\WaveFunction^\text{L}_{-}(\Pos, \TimeMax)}{\WaveFunction^\text{R}_{-}(\Pos)}
\end{pmatrix},
\end{align}
where $\WaveFunction^\text{L}_{\pm}(\Pos, \Time)$ is evolved according to the potential~\eqref{twowells}, we can define the state transfer fidelity,
\begin{align}
\Fidelity_\text{st} = \frac{1}{6}[ \Trac \ ( \AuxM' \AuxM'^{\dagger} ) + |  \Trac \ \AuxM' |^2 ],    
\end{align}
as the reward function.
In Fig.~\ref{fig5}(d,e), we show the example of a state of the two lowest vibrational levels of the left well transferred to the right one with $\Fidelity_\text{st} \approx 99\%$.

\subsection{Single-shot discrimination}\label{prot:state_meas}
\begin{figure}[h!t!]
    \centering
    \includegraphics[width=\linewidth]{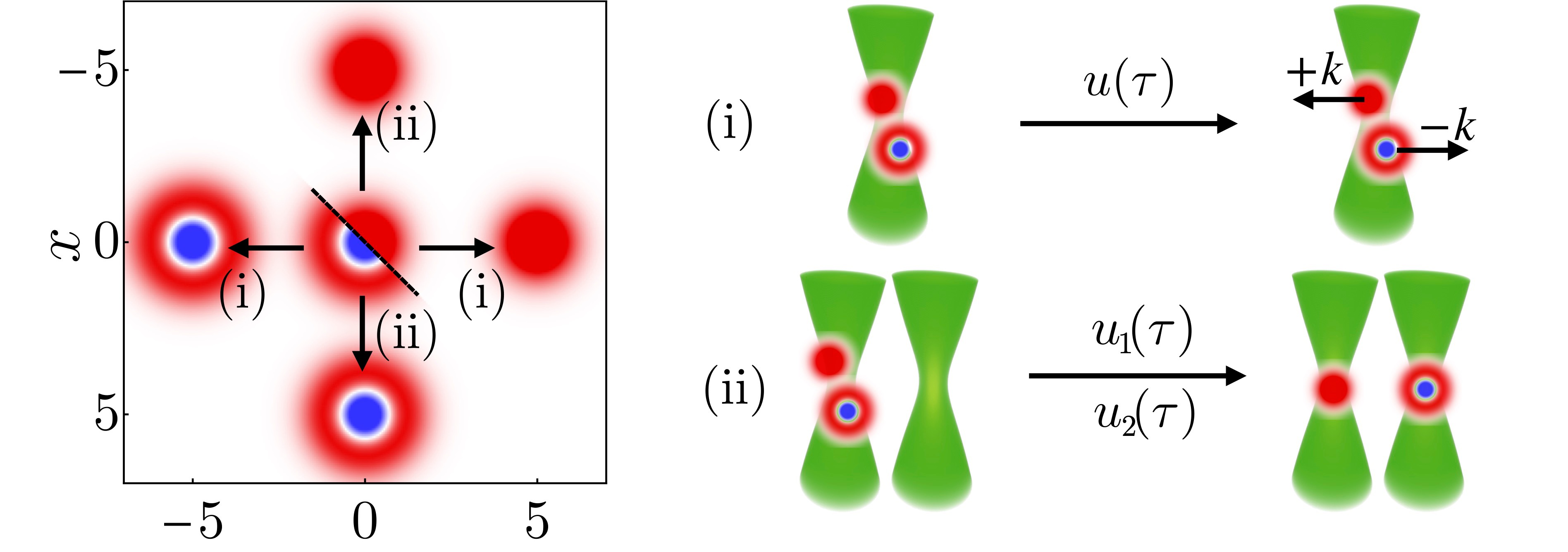}
    \caption{Two proposed discrimination protocols, based on a phase-space separation. (i) The opposite momentum kicks are imprinted onto each of the states. (ii) States are separated spatially via a second potential well.
    }
    \label{fig7}
\end{figure}

After presenting state preparation and unitary implementation in various nonharmonic potential landscapes, we move on to two protocols that perform single-shot discrimination between two orthogonal states $\WaveFunction^{\pm} (\Pos)$ (see Fig.~\ref{fig7}), providing an alternative to already existing methods involving auxiliary systems in, e.g., superconducting circuits~\cite{Schuster2007}.
The first one involves a single nonharmonic well and is performed through imprinting opposite momentum kicks onto each of the states via a potential displacement,
\begin{align}
\WaveFunction^{\pm}(\Pos) \rightarrow \WaveFunction^{\pm}_k(\Pos) =\WaveFunction^{\pm}(\Pos) e^{\pm i \MomKick \Pos},
\end{align}
where $\MomKick$ is chosen such that two phase-imprinted states are nonoverlapping in phase space.
After the phase imprinting, a selective measurement takes place, which can be realized in, e.g., an optical tweezer setup through the subsequent release of the trap, in which time-of-flight evolution reveals spatially separated detection clicks for each of the states~\cite{Fuhrmanek2010,Brown2023a}.
The reward function for such a momentum-kick protocol involves a sum of equally weighted fidelities,
\begin{align}\label{mkickfid}
\Fidelity_\text{mk} = \frac{1}{2}\sum_{j=\pm} \left|\braket{\WaveFunction^{j}(\Pos,\TimeMax)}{\WaveFunction^{j}_k(\Pos)} \right|^2  
\end{align}
\begin{figure}[h!t!]
    \includegraphics[width=\linewidth]{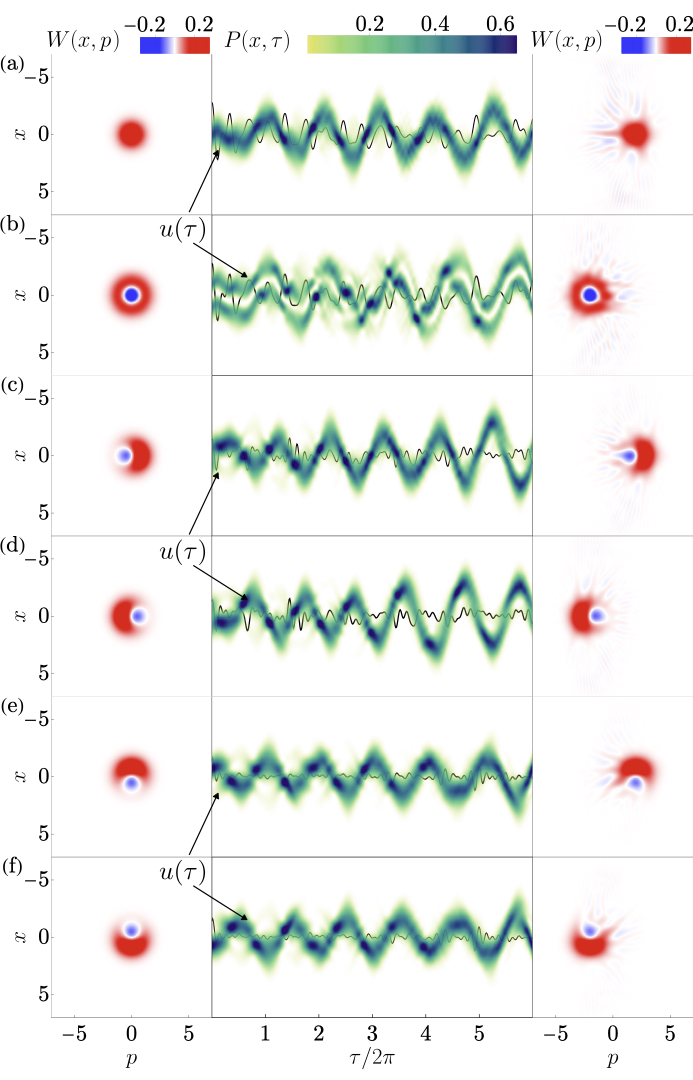}
    \caption{
    Examples of the implementation of momentum kick protocols via an optimally controlled single-well Gaussian potential with $\NonLin = 0.25$, $\TimeMax/ 2 \pi = 6$, and $\MomKick = 2$.
    (a-b) $\WaveFunction^{+}(\Pos) = \GroundState(\Pos)$ and $\WaveFunction^{-}(\Pos) = \WaveFunction_1(\Pos)$, (c-d) $\WaveFunction^{\pm}(\Pos) = [\GroundState(\Pos) \pm \WaveFunction_1(\Pos)]/\sqrt{2}$, and (e-f) $\WaveFunction^{\pm}(\Pos) = [\GroundState(\Pos) \pm i \WaveFunction_1(\Pos)]/\sqrt{2}$.
    Momentum-kick fidelities read (a-b) $\Fidelity_\text{mk} \approx 98.5 \%$, (c-d) $\Fidelity_\text{mk} \approx 99.4 \%$, and (e-f) $\Fidelity_\text{mk} \approx 99.5 \%$. See the caption of Fig.~\ref{fig1} for subplot and curve legend details. \label{fig8}}
\end{figure}
Note that in contrast to the unitary implementation, here the final relative phase between the states does not matter, as we aim only to distinguish the initial states.
As an example, in Fig.~\ref{fig7} $\WaveFunction^{\pm}(\Pos)$ are taken to be the two lowest eigenstates of the well, and the protocol is shown in a schematic way.
In Fig.~\ref{fig8}, we show in detail specific realizations of this and other momentum-kick protocols, utilizing a double-well Gaussian potential.
They also involve other choices of orthogonal states for the discrimination, namely eigenstates of both $\SigmaX$ and $\SigmaY$ Pauli matrices.
Such discrimination procedures then correspond to $\SigmaX$ and $\SigmaY$ measurements within a given two-level subspace.

The second discrimination protocol involves a double-well potential landscape and utilizes spatial separation instead of the momentum one (see Fig.~\ref{fig7}).
Initially, the states are fully confined to the left well and their discrimination is realized through a selective stealing protocol---if the state is $\WaveFunction^{+}(\Pos)$, then it ends up in the second well after the optimal shaking, while $\WaveFunction^{-}(\Pos)$ stays in the initial one. 
Then, a large spatial separation between the wells enables selective measurement.
The reward function for this protocol, the selective-stealing fidelity, reads
\begin{align}\label{ssfid}
\Fidelity_\text{ss} = & \frac{1}{2} \Big[ \left|\braket{\WaveFunction_{-}^\text{L}(\Pos,\TimeMax)}{\WaveFunction_{-}^\text{L}(\Pos)} \right|^2 \nonumber \\
& \quad \quad +\left|\braket{\WaveFunction_{+}^\text{L}(\Pos,\TimeMax)}{\WaveFunction_{+}^\text{R}(\Pos)} \right|^2 \Big], 
\end{align}
where the definitions~\eqref{leftwell} and~\eqref{righttwell} are used.
We show the exemplary case of distinguishing two lowest vibrational states in the two top panels of Fig.~\ref{fig7}, while further examples, including discrimination of $\SigmaX$ and $\SigmaY$ eigenstates, are presented in Appendix~\ref{app:Stealing}.

\subsection{Cooling}\label{prot:cooling}
These methods performed on low-lying eigenstates of a nonharmonic well can be straightforwardly extended into cooling protocols, similar to algorithmic~\cite{Popp2006,Shaw2025} or evaporative cooling.
These protocols involve bringing not only the first excited state $\FockStateOne(\Pos)$ but also higher excited states $\FockStateN(\Pos)$ into a second potential well or kicking them out of the well, followed by the destructive measurement of the separated excited fraction.
In the former case, the reward function reads
\begin{align}\label{sscoolfid1}
\Fidelity_\text{ss}^\text{cool} = & \frac{1}{N+1}  \Big[ \left|\braket{\WaveFunction_{0}^\text{L}(\Pos,\TimeMax)}{\WaveFunction_{0}^\text{L}(\Pos)} \right|^2 \nonumber \\
& \quad \quad + \sum_{n=1}^{N} \left|\braket{\WaveFunction_{n}^\text{L}(\Pos,\TimeMax)}{\WaveFunction_{n}^\text{R}(\Pos)} \right|^2 \Big], 
\end{align}
where 
\begin{align}
\WaveFunction_{n}^\text{L}(\Pos) =  \WaveFunction_{n}(\Pos + \TweezerSeparation/2), \nonumber \\
\WaveFunction_{n}^\text{R}(\Pos) =  \WaveFunction_{n}(\Pos - \TweezerSeparation/2).
\end{align}
\begin{figure}[h!t!]
    \centering
    \includegraphics[width=\linewidth]{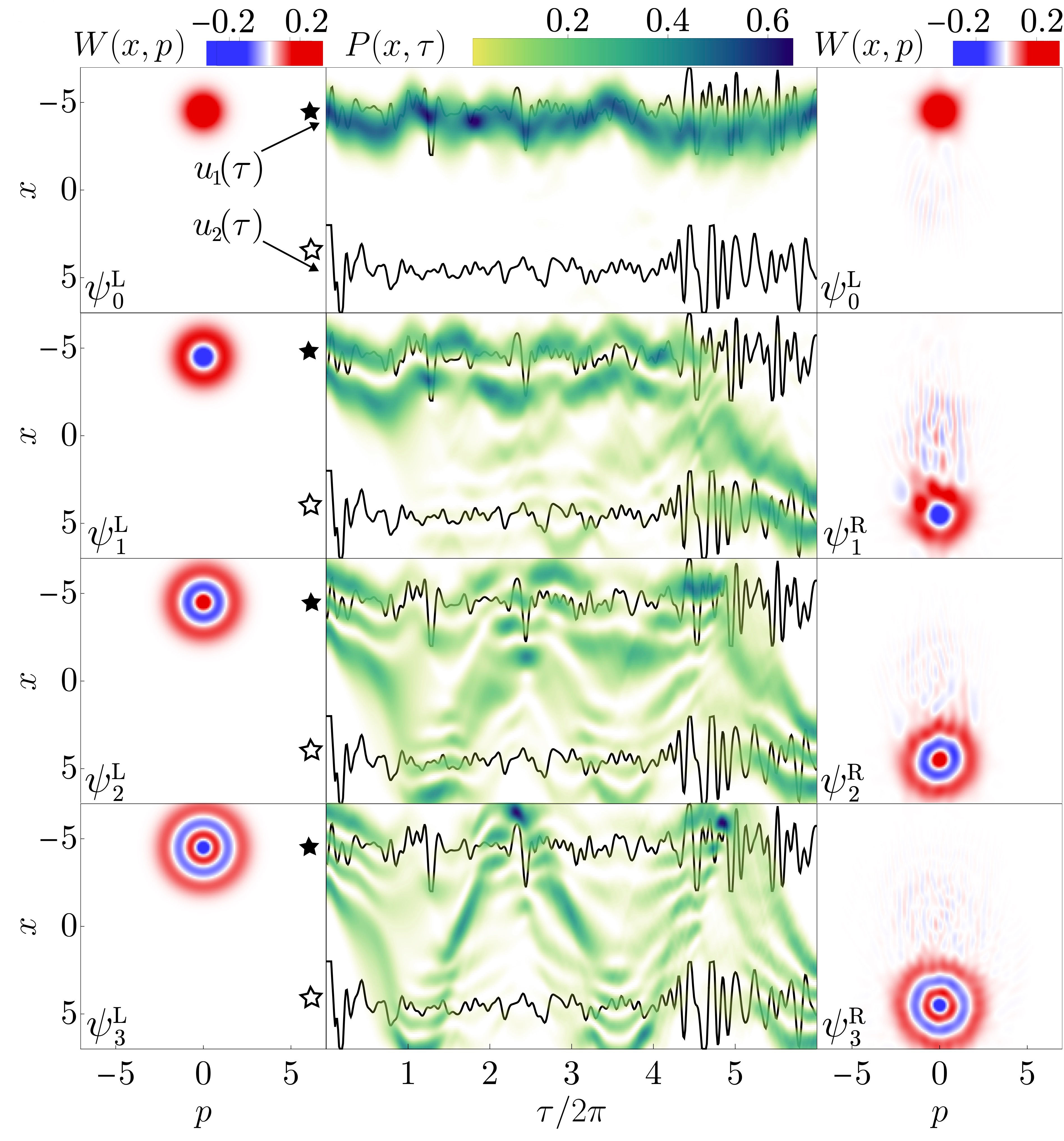}
    \caption{The accelerated algorithmic cooling method by spatial separation of the ground and excited fractions realized through optimal control of two wells. 
    The selective-stealing cooling fidelity reads $\Fidelity_\text{ss}^\text{cool} \approx 97 \%.$ See the caption of Fig.~\ref{fig1} for subplot and curve legend details.
    }
    \label{fig9}
\end{figure}
Here, $N$ characterizes the number of excited states that are brought to the other well.
The higher $N$, the more thermal state can be brought to the ground state.
However, higher temperature also means a lower chance of success, as the presence of the ground state in the left well is conditioned on the subsequent destructive measurement in the right well.
Hence, on average, the success rate of the protocol is proportional to the initial occupation of the ground state.
Fig.~\ref{fig9} demonstrates this process for the four lowest states of the well, with the higher three states becoming spatially separated from the ground state.
The method involving the momentum-kick protocol is analogous, but with the excited fraction being kicked out of the well.
The reward function is thus
\begin{align}\label{sscoolfid2}
\Fidelity_\text{mk}^\text{cool} = & \frac{1}{N+1}  \Big[ \left|\braket{\WaveFunction_{0}(\Pos,\TimeMax)}{\WaveFunction_{0}(\Pos)} \right|^2 \nonumber \\
& \quad \quad + \sum_{n=1}^{N} \left|\braket{\WaveFunction_{n}(\Pos,\TimeMax)}{\WaveFunction_{n}^k(\Pos)} \right|^2 \Big], 
\end{align}
where
\begin{align}\label{kickedFock}
\WaveFunction_{n}^k(\Pos) = e^{i k x}\WaveFunction_{n}(\Pos).
\end{align}

\section{Feasibility}\label{sec:feasibility}
After presenting several protocols, let us now address aspects relevant to their experimental implementation.
From now on, we will focus on the Fock state preparation protocol with a single-well Gaussian potential as an exemplary case.
In subsection~\ref{qsl}, we analyze how fast the optimized protocol can be and how excited the system is during its evolution. 
Then, in subsection~\ref{noise}, we check the robustness of the state preparation in the presence of noise and decoherence modeled by the stochastic fluctuations of the potential.
Finally, subsection~\ref{couplings} addresses the issue of satisfying one-dimensional approximation in the specific case of an atom in an optical tweezer.

\subsection{Quantum speed limit}\label{qsl}
\begin{figure}[ht!]
    \centering
    \includegraphics[width=\linewidth]{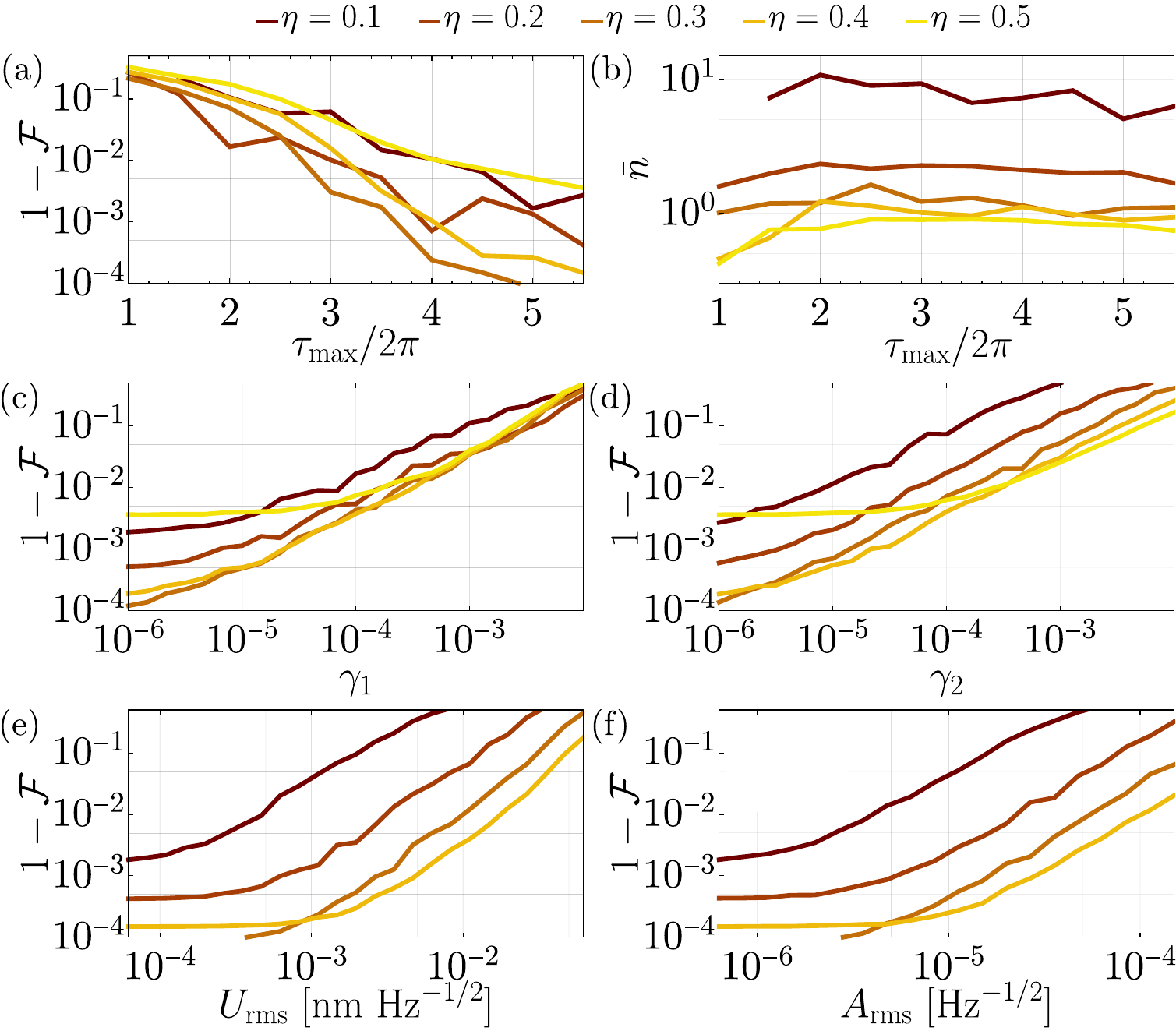}
\caption{ (a) Infidelity of the ground to the first excited state transfer, $\GroundState(\Pos) \rightarrow \FockStateOne(\Pos)$, for a Gaussian potential as a function of the length of protocol $\TimeMax$ and nonharmonicity $\NonLin$.
The scaling of the curve is as expected from the previous studies~\cite{vanFrank2016} and reveals that a quantum speed limit $\QSL$ does not scale strongly with $\NonLin$.
(b) Average excitation number as a function of $\TimeMax$ and $\NonLin$.
(c,d) The minimal achievable infidelity for given values of position ($\gamma_1$) and depth ($\gamma_2$) noises for a Gaussian potential.
(e,f) The minimal achievable infidelity for given values of displacement $\RMSPointing$ and intensity $\RMSIntensity$ noises for a rubidium-87 atom in an optical tweezer with the waist $\TweezerWaist = 710 \ $nm.
The value of nonharmonicity $\NonLin$ is tuned via the tweezer's depth $\TweezerDepth$.
    }
    \label{fig10}
\end{figure}

Concerning the protocol's speed, a fundamental restriction exists, the quantum speed limit~\cite{Deffner2017}, which gives the minimal time that is needed to perform a specific state transfer with a given fidelity threshold.
Well-understood for two-level systems driven by static Hamiltonians~\cite{Deffner2017,Mandelstam1945,Margolus1998}, quantum speed limits have also been analyzed for time-dependent multilevel systems~\cite{Anandan1990,Deffner2013}, and, specifically, for the ground to the first excited state transfer in a Bose-Einstein condensate~\cite{vanFrank2016}.
We analyze a Gaussian potential case to find a minimum time $\QSL$ needed to perform the excitation from the ground to the first excited state, $\GroundState(\Pos) \rightarrow \FockStateOne(\Pos)$, with some infidelity threshold for different values of nonharmonicity $\NonLin$ .
In the regime of interest, $\NonLin \sim 0.1 - 0.5$, there is no strong scaling of $\QSL$ with the nonharmonicity and achievable fidelity scales exponentially with the protocol time.
For the example of an optical tweezer, it implies that, without noise, the deeper the tweezer, the more bound states accessed and the faster the protocol (and hence requiring more control bandwidth), as $\Frequency \sim \TweezerDepth^{1/2}$.
We then analyze the number of excitations averaged throughout the protocol,
\begin{align}
\AvN = \int d \Time \sum_n n \OccupNumbers,
\end{align}
as this quantity is relevant for analytical quantum speed limits~\cite{Margolus1998,Chen2010a,Ness2021} and sensitivity to decoherence.  
Fig.~\ref{fig10}(b) shows that the average excitation number increases as the nonharmonicity decreases.
This indicates that protocols involving deeper, more harmonic potential wells tend to populate higher excited states.
Interestingly, $\AvN$ appears to depend only weakly on $\TimeMax$.
A plausible explanation is that the maximum excitation is ultimately constrained by the finite depth of the potential and is saturated in optimally controlled scenarios.
Consequently, highly nonharmonic potentials are preferred when minimizing excitation (and thus enhancing robustness) is important.

\subsection{Noise and decoherence}\label{noise}
Let us now address the noise robustness of the state preparation protocols.
In general, for different experimental realizations, typical sources of error include stochastic vibrations of the potential, low-frequency drifts between experimental runs, imperfections of the control, etc.
To model potential vibrations and control imperfections in a single potential well, we employ stochastic fluctuations of the depth and position of the potential, 
\begin{align}
    \InControl(\Time) &\rightarrow \InControl(\Time) + \Fluct_2(\Time), \nonumber \\
    \PosControl(\Time) &\rightarrow \PosControl(\Time) + \Fluct_1(\Time),
\end{align}
where $\Fluct_j(\Time)$ for $j=1,2$ are dimensionless stochastic Gaussian variables of zero mean and assumed delta-correlated in the relevant time scales, namely
\begin{align}
\left\langle \Fluct_j(\Time) \Fluct_j(\Time') \right\rangle =  \Noise_j \delta(\Time-\Time').
\end{align}
We solve noisy dynamics by averaging over different noise realizations and constructing a Monte Carlo density matrix~\cite{Molmer1993}.
In Figs.~\ref{fig10}(c,d), we plot the minimal achievable infidelity at a given level of noise for the ground to the first excited state excitation protocol for a single Gaussian potential.
While the positional noise affects the protocol comparably for each $\NonLin$, the depth noise influences the strongest the least nonharmonic case.
For the specific mapping onto pointing and intensity noise in an optical tweezer setup, we have 
\begin{align}
\Noise_1 &= \frac{\RMSPointing^2 \Frequency}{\OscillatorLength^2}, \nonumber \\
\Noise_2 &= \RMSIntensity^2 \Frequency,
\end{align}
where $\RMSPointing$ is a pointing root mean square (RMS) noise, and $\RMSIntensity$ is an intensity RMS noise.
In Figs.~\ref{fig10}(e,f), we plot the fidelity loss as a function of $\RMSPointing$ and $\RMSIntensity$, and we find that shallower (more nonharmonic) tweezers outperform the deeper ones.

Let us compare this observation with the charge noise in superconducting transmon qubits.
In the transmon regime, the dephasing rate due to charge noise is exponentially suppressed with decreasing nonharmonicity, following $\Noise \propto \exp\small(-\sqrt{8 \JosEnergyTot / \ChargeEnergy}\small) = \exp\small(-2 / \NonLin^2\small)$~\cite{Koch2007}.
This trend is opposite to what we observe in Figs.~\ref{fig10}(c,d).
There, the fidelity increases with increasing $\NonLin$, but the dependence is at most polynomial.
This suggests that, unlike optically trapped atoms, transmons benefit from less nonharmonic potentials.

\subsection{Coherent couplings}\label{couplings}
Coherent coupling to other degrees of freedom might cause an additional experimental challenge.
The exemplary case of such a noisy channel occurs in one of the setups we consider and is due to the dimensionality of the realistic optical potential.
The three-dimensional tweezer potential is necessarily nonseparable and, depending on the axis of control, axial or radial, a one-dimensional approximation may be less strictly satisfied~\cite{Mennemann2015}.
Restricting control to only one direction unavoidably leads to excitation in other ones.
However, the fidelity loss due to these excitations can be alleviated through additional optimization in a fully three-dimensional potential and additional optimized position and depth controls.
The additional control then effectively deexcites the coupled degrees of freedom.
This optimization can be done via a staged procedure, where the initial guess for a higher-dimensional system comes from a reduced-dimensional control optimization.
Such an optimization procedure can be utilized in setups in which coherent dynamics beyond one-dimensional approximation can be solved.

Here we present the analysis of the Fock $\FockStateOne(\Pos)$ state preparation protocol in the two- and three-dimensional geometry.
The dynamics is driven by either two-dimensional,
\begin{align}
    \DimPotNon_\text{2D} (\Pos,\PosY) &= \frac{3}{4 \NonLin^2} \left[ 1 - \exp( - \frac{2 }{3} \NonLin^2 (\Pos^2 + \PosY^2)  ) \right] \nonumber \\
    &\approx \frac{1}{2} \Pos^2 + \frac{1}{2} \PosY^2,
\end{align}
or three-dimensional Gaussian potential,
\begin{figure}[ht!]
    \includegraphics[width=\linewidth]{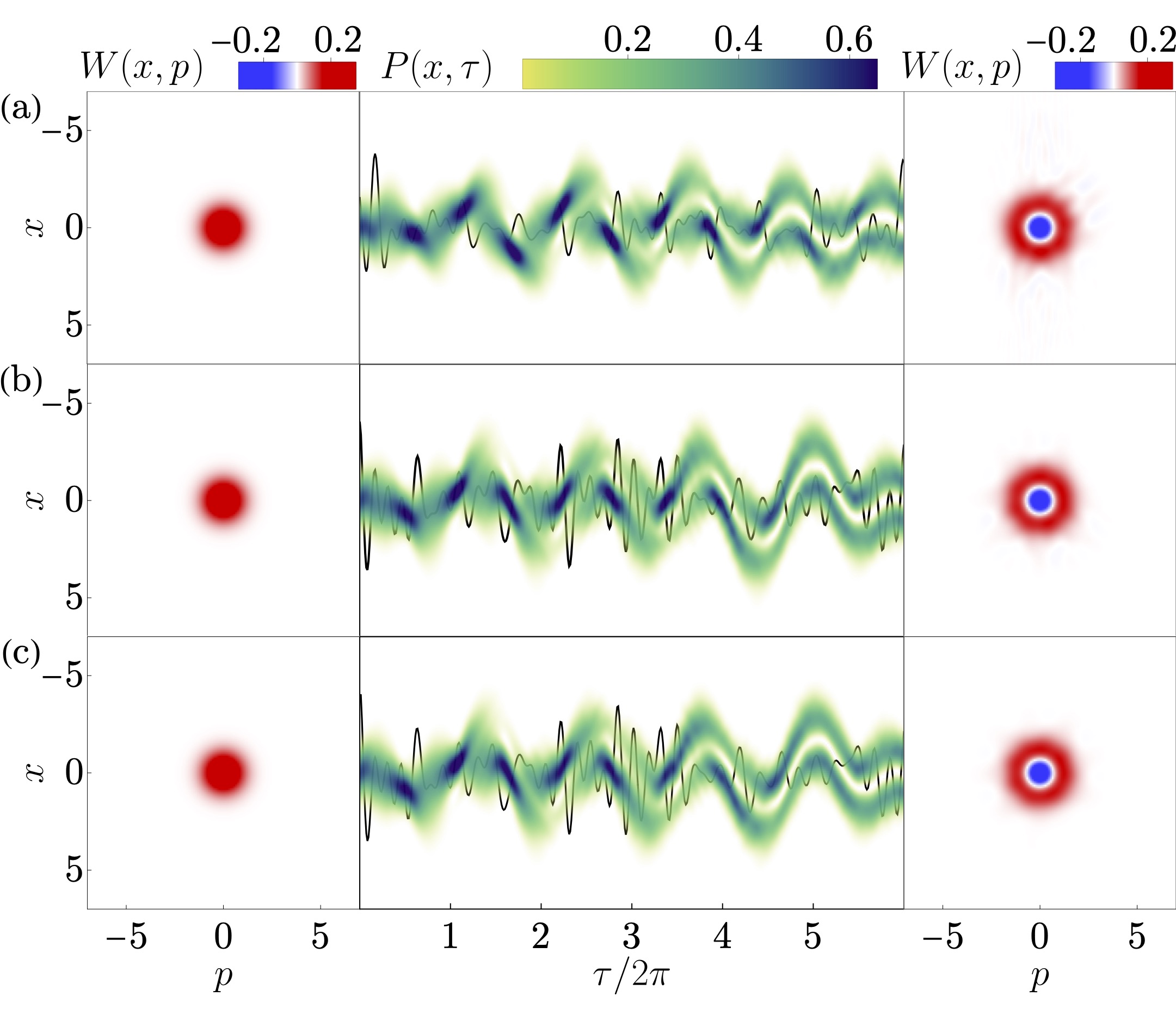}
    \caption{ Fock $\FockStateOne(\Pos)$ state preparation protocol with the optimally controlled Gaussian potential characterized with $\NonLin = 0.25$ and $\AspectRatio = 0.167$ for (a) one-, (b) two-, and (c) three-dimensional geometries.
    Achieved fidelities $\Fidelity$ read (a) $99.7 \%$, (b) $99.4 \%$, and (c) $99.0 \%$. See the caption of Fig.~\ref{fig1} for subplot and curve legend details.\label{fig11}}
\end{figure}
\begin{align}
    \DimPotNon_\text{3D} (\Pos,\PosY,\PosZ) &= \frac{3}{4 \NonLin^2} \frac{  1+\frac{2}{3} \NonLin^2 \AspectRatio^2 \PosZ^2 -  \exp[ - \frac{2 }{3} \frac{ \NonLin^2 (\Pos^2 + \PosY^2)}{1+\frac{2}{3} \NonLin^2 \AspectRatio^2 \PosZ^2}  ]}{1+\frac{2}{3} \NonLin^2 \AspectRatio^2 \PosZ^2}, \nonumber \\
    &\approx \frac{1}{2} \Pos^2 + \frac{1}{2} \PosY^2 + \frac{1}{2} \AspectRatio^2 \PosZ^2 
\end{align}
where $\AspectRatio$ is the aspect ratio between the radial and axial trapping frequencies, $\AspectRatio = \Frequency / \FreqAxial$, and the leading harmonic approximations are given.
The initial states are taken to be the ground states of the harmonic approximations,
\begin{align}
\GroundState(\Pos,\PosY)   &= \GroundState(\Pos) \GroundState(\PosY), \nonumber \\
\GroundState(\Pos,\PosY,\PosZ)   &= \GroundState(\Pos) \GroundState(\PosY) \AspectRatio^{1/4} \GroundState(\AspectRatio\PosZ).
\end{align}
The control $\PosControl(\Time)$ is taken to be only $x$-displacement, 
\begin{align}
\DimPotNon_\text{2D} (\Pos,\PosY,\Time)   &= \DimPotNon_\text{2D} [\Pos - \PosControl(\Time),\PosY] , \nonumber \\
\DimPotNon_\text{3D} (\Pos,\PosY,\PosZ,\Time)    &= \DimPotNon_\text{3D} [\Pos - \PosControl(\Time),\PosY,\PosZ],
\end{align}
and optimized through a staged procedure: the optimized control in the one-dimensional problem is a starting guess for the two-dimensional control; analogously for the three-dimensional case.
In Fig.~\ref{fig11}, we plot the results of such an optimization with the snapshots of Wigner functions associated with reduced density matrices,
\begin{align}
\DensityMatrix_{\text{2D}} &= \Tr_{y} \DensityMatrix, \nonumber \\
\DensityMatrix_{\text{3D}} &= \Tr_{y,z} \DensityMatrix, 
\end{align}
for two- and three-dimensional cases, respectively.
Here, $\DensityMatrix$ is the full motional density matrix.
As for the low-frequency drifts, one can further optimize the control functions in the presence of varying experimental parameters, a commonly utilized technique~\cite{Rossignolo2023}.
Going beyond the single-well case, note that arbitrary control of double-well systems can lead to additional sources of noise, including, e.g., relative bias fluctuations.

\section{Conclusions \& outlook}\label{sec:conclusions}

In conclusion, we have presented an approach that allows the preparation and manipulation of a state of a single continuous-variable degree of freedom and relies only on self-contained nonlinear dynamics without any additional nonlinear couplings.
Specifically, the presented method utilizes optimal control of the intrinsic nonharmonicity present in the effective confining potential, a common scenario in many continuous-variable systems, including neutral atoms in optical tweezers and lattices or flux-tunable transmons.
We have presented quantum state preparation, implementation of arbitrary unitaries within a selected subspace, quantum discrimination protocols based on phase-space separation, algorithmic cooling, spatial logical state transfer, and protection from state deformation during nonlinear dynamics that can be realized in such systems.
We have shown that these protocols can be performed via both single- and double-well potential landscapes.
The protocols that we have proposed are particularly well-suited for systems in which coherent couplings to two-level degrees of freedom have not been realized.
Beyond such systems, they can be implemented in setups with already well-developed quantum control, such as superconducting circuits, either as a proof of principle, as an addition to an already existing experimental toolbox, or as a way to unify the utilization of different types of quantum non-Gaussian states.

Moreover, we have analyzed the performance of the state preparation protocols, from the perspective of the speed of the protocols, under the effect of position and depth noises, and with additional coherent couplings.
These aspects have been checked for a specific example of an atom in an optical tweezer.
Our proposal is compatible with state-of-the-art technology, such as neutral atoms, circuit quantum electrodynamics, or Bose-Einstein condensates, and might be explored in systems with smaller nonharmonicities, e.g.,~levitated nanoparticles ($\NonLin \sim 10^{-5}$)~\cite{Gonzalez-Ballestero2021,Roda-Llordes2024b} or ions in Paul traps (typically $\NonLin \sim 10^{-2} - 10^{-1}$, which allows a crossover between weak and strong nonharmonicity)~\cite{Leibfried2003,Home2011}.
More specifically, for the tasks we consider, controllability decreases with decreasing noharmonicity, confirming notorious difficulty in preparing non-Gaussian states in weakly nonharmonic systems.
The strategies to circumvent this problem involve large excitation of involved states through, e.g., phase-displacement in cavities for Fock state preparation~\cite{Lingenfelter2021,Yuan2025} or motional squeezing in mechanical oscillators for cubic-~\cite{Roda-Llordes2024b,Roda-Llordes2024a,Riera-Campeny2024} and quartic-phase state preparation~\cite{Rosiek2024}.
Combining optimal control of the potential landscape with such a nonharmonicity enhancement may bring these weakly nonharmonic systems closer to universal quantum control in a resource-optimal way.

\section*{Data and materials availability}
The Fourier transforms of all presented controls are presented in Appendix~\ref{fourier_pulses}, while the depth modulations not shown in the main text are plotted in Appendix~\ref{depth}.
Data analysis and simulation codes are available on Zenodo~\cite{zenodo}.

\begin{acknowledgments}
We thank R. G. Corti{\~n}as, N. E. Frattini, S. Muleady, A. M. Rey, and P. Zoller for helpful discussions.
The computational results presented here have been achieved in part using the LEO HPC infrastructure of the University of Innsbruck.
P.T.G. and O.R.-I. have been supported by the European Union’s Horizon 2020 research and innovation programme under grant agreement No. 863132 (IQLev) and by the European Research Council (ERC) under the grant agreement No. 951234 (Q-Xtreme ERC-2020-SyG).
P.T.G. was partially supported by the project CZ.02.01.01/00/22\_008/0004649 (QUEENTEC) of EU and project 23-06308S of the Czech Science Foundation.
H.P. has been supported through an ERC Starting grant QARA (grant no.~101041435) and the European Union's Horizon 2020 research and innovation program under Grant Agreement No. 101079862 (PASQuanS2).
C.A.R. has been supported by NSF PHY-2317149, NSF QLCI award OMA 2016244, the US Department of Energy, Office of Science, National Quantum Information Science Research Centers, Quantum Systems Accelerator, and the Baur-SPIE Chair at JILA.

\end{acknowledgments}

%\bibliographystyle{quantum}
% \bibliography{ShakingPaper}
\bibliographystyle{apsrev4-1}
\setlength{\bibsep}{0pt plus 0.3ex}
%\nocite{apsrev41Control}
\bibliography{ShakingPaper,revtex-custom}

\newpage

\appendix

\renewcommand{\thefigure}{A\arabic{figure}}
\setcounter{figure}{0}

\section{Single-well two-level unitary implementations\label{app:gates1}}
\begin{figure}[H]
    \includegraphics[width=\linewidth]{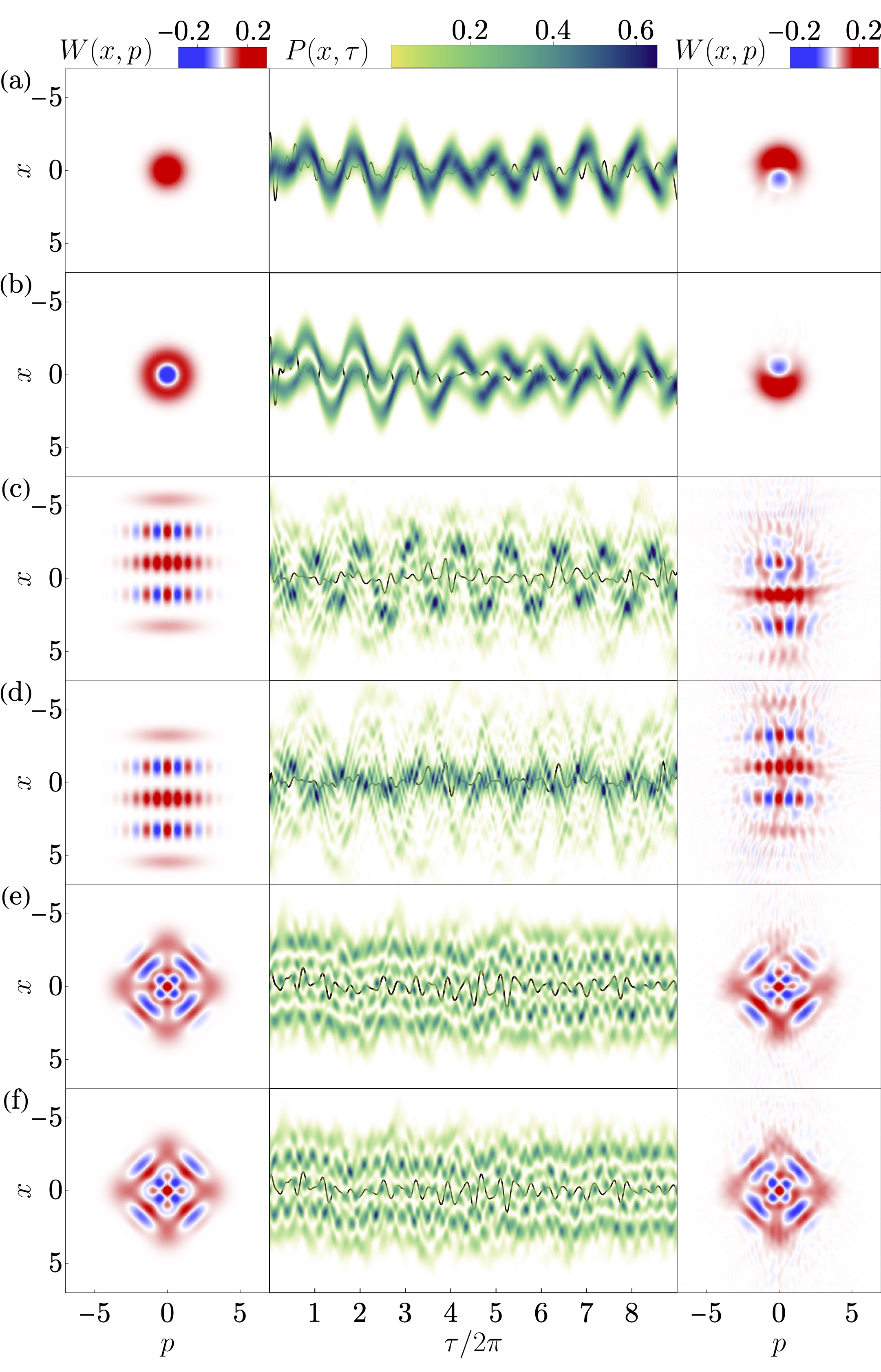}
    \caption{Evolution of selected orthogonal states in a cosine potential characterized by $\NonLin = 0.2$ and with optimally controlled displacement $\PosControl(\Time)$ and intensity $\InControl(\Time)$ that are constrained through the flux control of a flux-tunable transmon with $\JunctionAssymetry = 0.8$.
    The solid line shows $\PosControl(\Time)$ and the protocol lasts $\TimeMax/ 2 \pi = 9$.
    (a-b) Hadamard unitary within a subspace spanned by $\ket{0}$ and $\ket{1}$ Fock states.
    The fidelity reads $\FidelityGate \approx 99.99 \%$.
    (c-d) $\SigmaY$ unitary within a subspace spanned by GKP states, $\WaveFunction^{\pm} = \GKP(\Pos \pm \Disp_3/4)$ with $\Squeeze = 0.7$, $(\Disp_1,\Disp_2,\Disp_3) = (-\sqrt{6 \pi},0,\sqrt{6 \pi})$.
    The fidelity reads $\FidelityGate \approx 89.5 \%$.
    (e-f) $\SigmaX$ unitary within a subspace spanned by four-legged-cat states states with $\CohDis = 2$.
    The fidelity reads $\FidelityGate \approx 94.0 \%$. See the caption of Fig.~\ref{fig1} for subplot and curve legend details.\label{figA1}}
\end{figure}

In this section, we present examples of the implementation of selected unitaries
within a two-level subspace spanned by either Fock, GKP, or four-legged-cat states.
Fig.~\ref{fig6} shows results for a single-well Gaussian potential, while Fig.~\ref{figA1} concerns a cosine one.

\newpage

\renewcommand{\thefigure}{B\arabic{figure}}
\setcounter{figure}{0}

\section{Double-well two-level unitary implementations\label{app:gates2}}
\begin{figure}[H]
    \includegraphics[width=\linewidth]{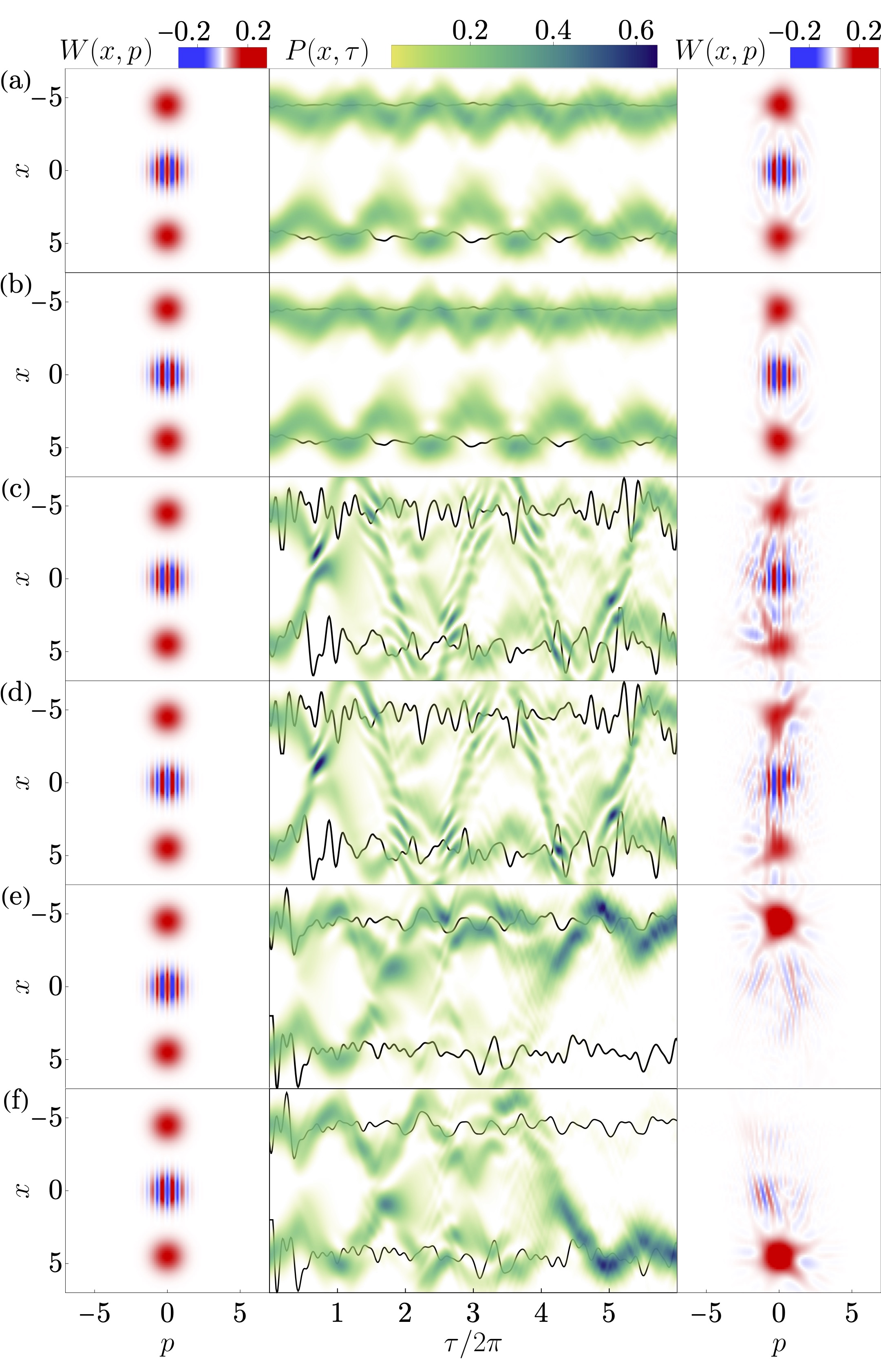}
    \caption{ Evolution of selected orthogonal Kerr-cat states in a double-well Gaussian potential with optimally controlled position displacements $\PosControl_1(\Time)$ and $\PosControl_2(\Time)$. 
    The solid lines show $\PosControl_1(\Time)$ and $\PosControl_2(\Time)$ and the protocol lasts $\TimeMax/ 2 \pi = 6$.
    (a-b) $\SigmaX$ unitary.
    The fidelity reads $\FidelityGate \approx 99.2 \%$.
    (c-d) $\SigmaX$ unitary.
    The fidelity reads $\FidelityGate \approx 90.9 \%$.
    (e-f) Hadamard unitary.
    The fidelity reads $\FidelityGate \approx 95.9 \%$. See the caption of Fig.~\ref{fig1} for subplot and curve legend details.\label{figB1}}
\end{figure}

In Fig.~\ref{figB1}, we present examples of the implementation of selected unitaries within a two-level subspace spanned by Kerr-cat states, $ \WaveFunction^\pm (\Pos) = [ \GroundState(\Pos + \TweezerSeparation / 2) \pm i \GroundState(\Pos - \TweezerSeparation / 2 )]/ \sqrt{2}$ with $\TweezerSeparation = 9$ via a double-well Gaussian potential with $\NonLin = 0.25$.

\newpage

\renewcommand{\thefigure}{C\arabic{figure}}
\setcounter{figure}{0}

\section{Selective stealing protocols\label{app:Stealing}}
\begin{figure}[H]
    \includegraphics[width=\linewidth]{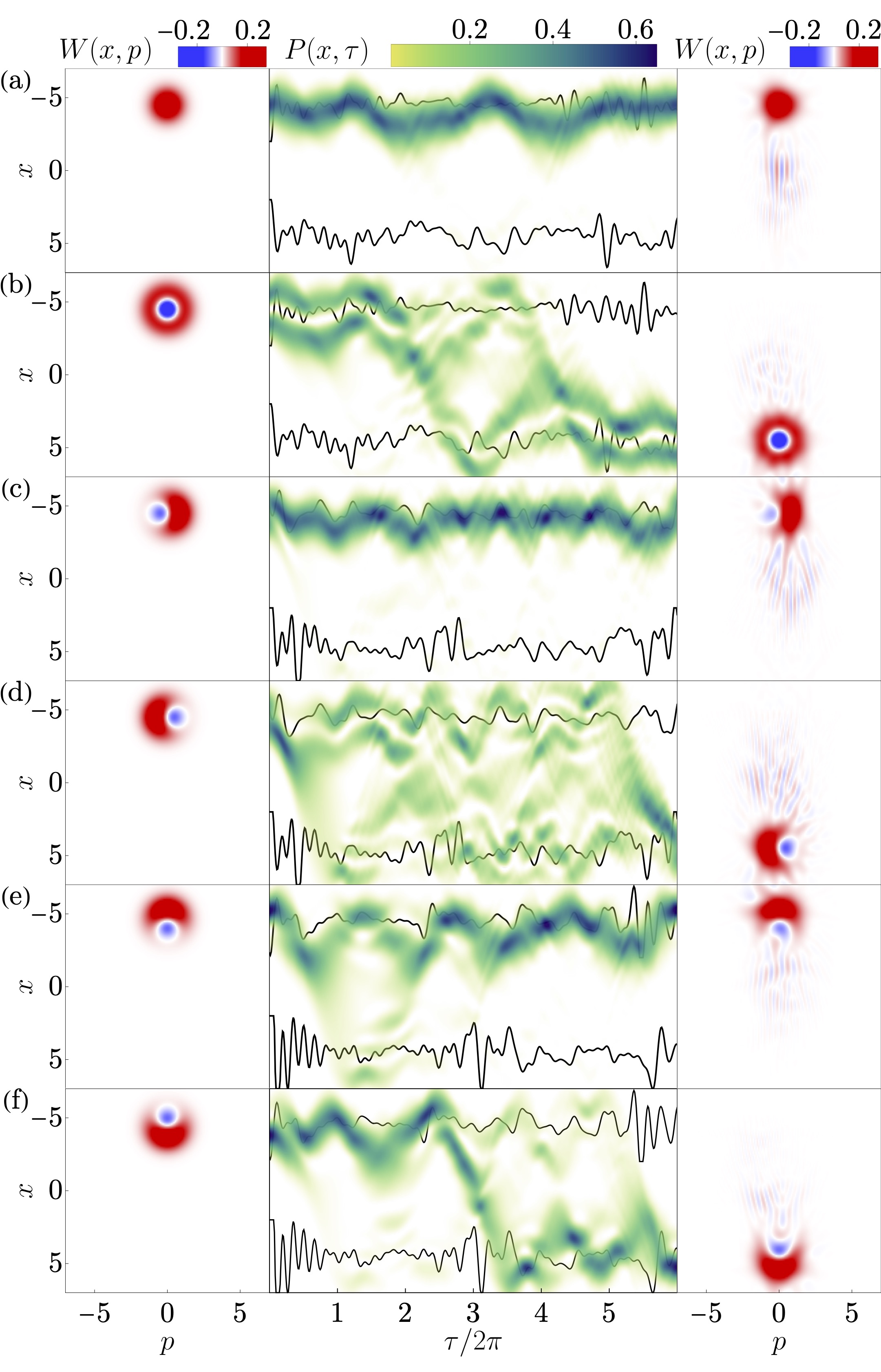}
    \caption{Spatial separation of orthogonal states $\WaveFunction^{\pm} (\Pos + \TweezerSeparation/2) \rightarrow \WaveFunction^{\pm} (\Pos \mp \TweezerSeparation/2)$ with $\TweezerSeparation = 9$ for (a-b) $\WaveFunction^{+} (\Pos)= \GroundState (\Pos)$ and $\WaveFunction^{-}(\Pos) = \FockStateOne (\Pos)$, (c-d) $\WaveFunction^{\pm} (\Pos)= [\GroundState (\Pos) \pm \FockStateOne (\Pos)]/\sqrt{2}$, and (e-f) $\WaveFunction^{\pm}  (\Pos) = [\GroundState (\Pos) \pm i \FockStateOne (\Pos)]/\sqrt{2}$.
    Fidelities read (a-b) $\FidelityGate \approx 98.9 \%$, (c-d) $\FidelityGate \approx 98.9 \%$, and (e-f) $\FidelityGate \approx 97.4 \%$. See the caption of Fig.~\ref{fig1} for subplot and curve legend details.\label{figC1}}
\end{figure}

In Fig.~\ref{figC1}, we present examples of the implementation of selective stealing protocols via an optimally controlled double-well Gaussian potential with $\NonLin = 0.25$ and $\TimeMax/ 2 \pi = 6$.

\newpage

\section{Fourier transforms of control functions\label{fourier_pulses}}
In this section, we present Fourier transforms of displacement control functions,
\begin{equation}
\mathfrak{F}_{\PosControl} = \left| \int \PosControl(\Time) \exp(-i\frac{\tilde{\Frequency}}{\Frequency} \Time)\right|,
\end{equation}
for all the presented protocols.
Specifically, Figs.~\ref{FigD-1},~\ref{FigD-5}, and~\ref{FigD-9} pertain to the figures from the main text, while Figs.~\ref{FigD-2},~\ref{FigD-6},~\ref{FigD-A1},~\ref{FigD-B1},~\ref{FigD-8},~\ref{FigD-C1}, and~\ref{FigD-11} are associated to the Supplemental Materials.

\renewcommand{\thefigure}{D-\arabic{figure}}
\setcounter{figure}{0}

\begin{figure}[h!]
    \includegraphics[width=0.97\linewidth]{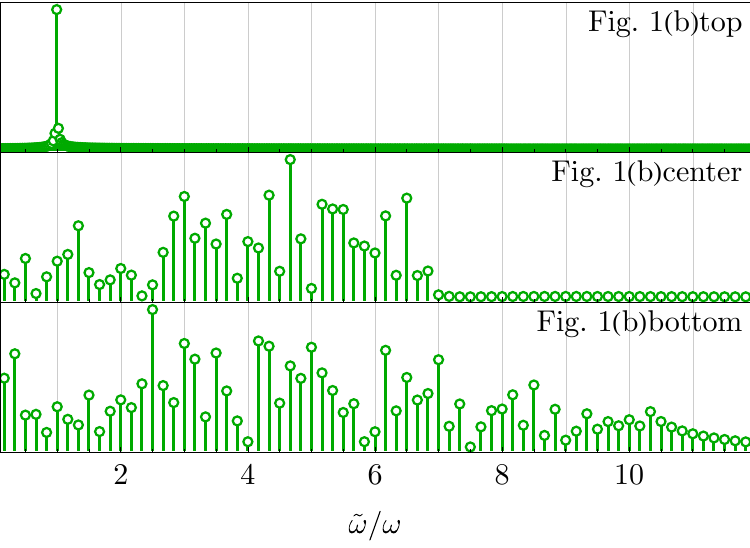}
    \caption{ Fourier transforms of displacement control functions $\PosControl(\Time)$ from Fig.~\ref{fig1} in the main text.
    The vertical axis is in arbitrary units, normalized to the highest frequency contribution.
    \label{FigD-1}}
\end{figure}

\newpage

\begin{figure}[h!]
    \includegraphics[width=0.97\linewidth]{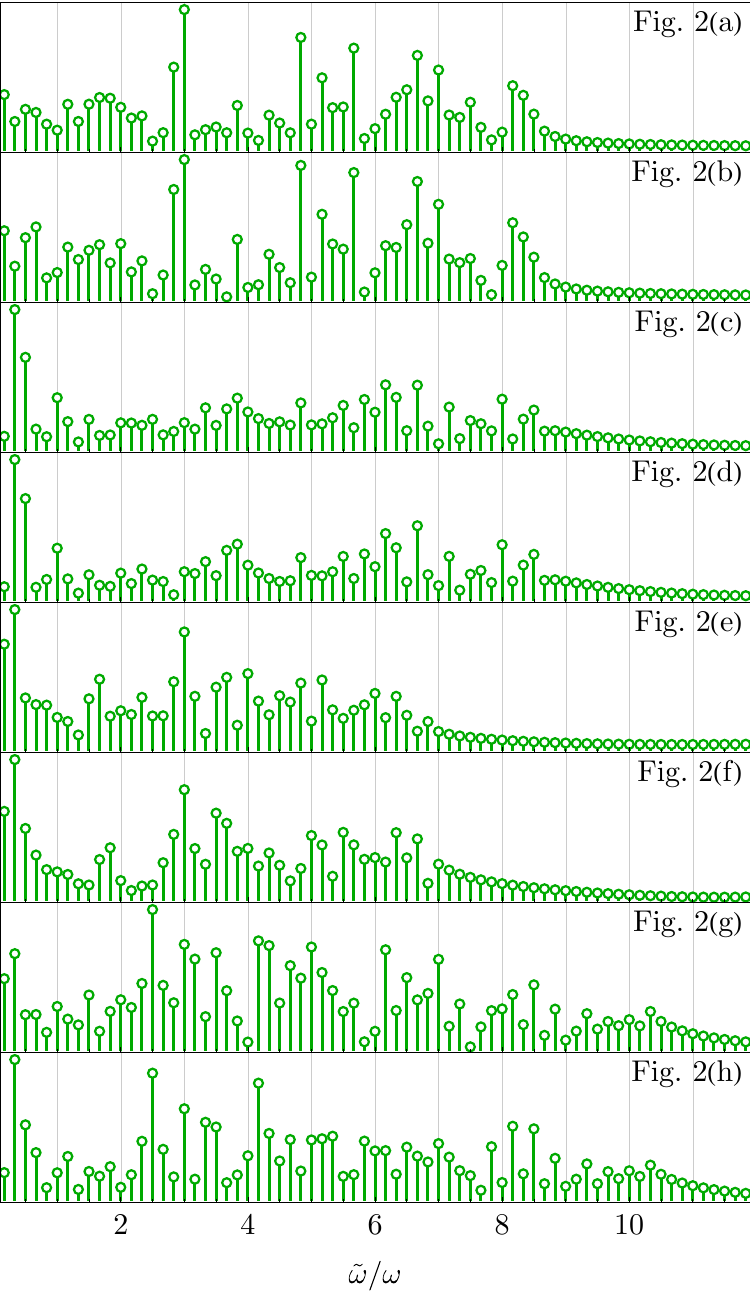}
    \caption{ Fourier transforms of displacement control functions $\PosControl(\Time)$ from Fig.~\ref{fig2}.
    The vertical axis is in arbitrary units, normalized to the highest frequency contribution.
    \label{FigD-2}}
\end{figure}

\begin{figure}[h!]
    \includegraphics[width=0.97\linewidth]{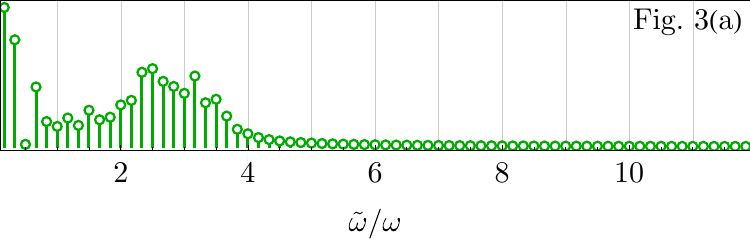}
    \caption{ Fourier transform of displacement control function $\PosControl(\Time)$ from Fig.~\ref{fig3}(a).
    The vertical axis is in arbitrary units, normalized to the highest frequency contribution.
    \label{FigD-3}}
\end{figure}

\newpage

\renewcommand{\thefigure}{D-\arabic{figure}}
\setcounter{figure}{4}

\begin{figure}[hb!]
    \includegraphics[width=0.97\linewidth]{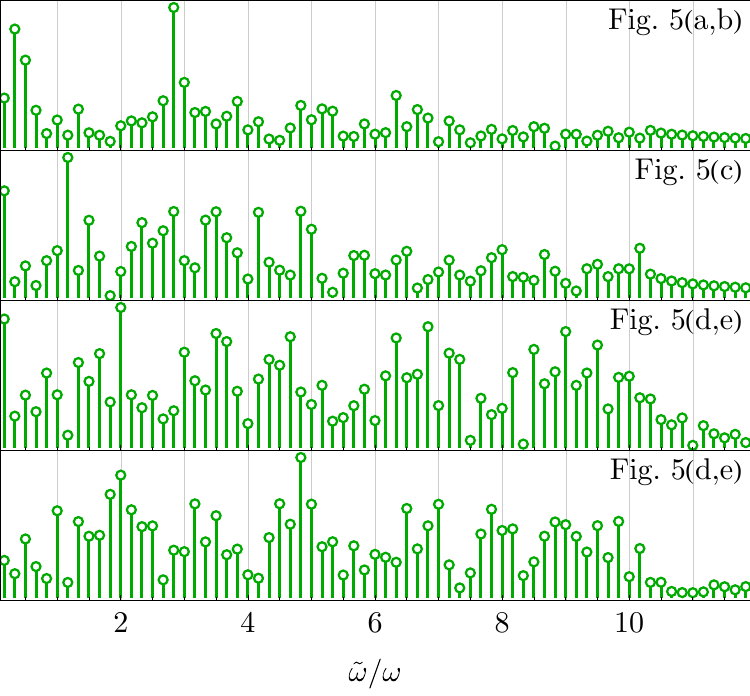}
    \caption{ Fourier transforms of displacement control functions $\PosControl(\Time)$ from Fig.~\ref{fig5} in the main text.
    The vertical axis is in arbitrary units, normalized to the highest frequency contribution.
    \label{FigD-5}}
\end{figure}

\begin{figure}[h!]
    \includegraphics[width=\linewidth]{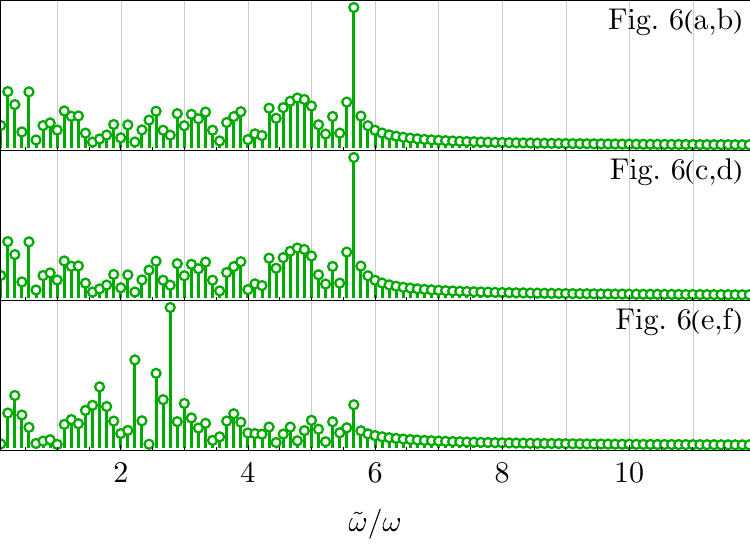}
    \caption{ Fourier transforms of displacement control functions $\PosControl(\Time)$ from Fig.~\ref{fig6}.
    The vertical axis is in arbitrary units, normalized to the highest frequency contribution.
    \label{FigD-6}}
\end{figure}

\newpage
\renewcommand{\thefigure}{D-\arabic{figure}}
\setcounter{figure}{7}

\begin{figure}[h!]
    \includegraphics[width=\linewidth]{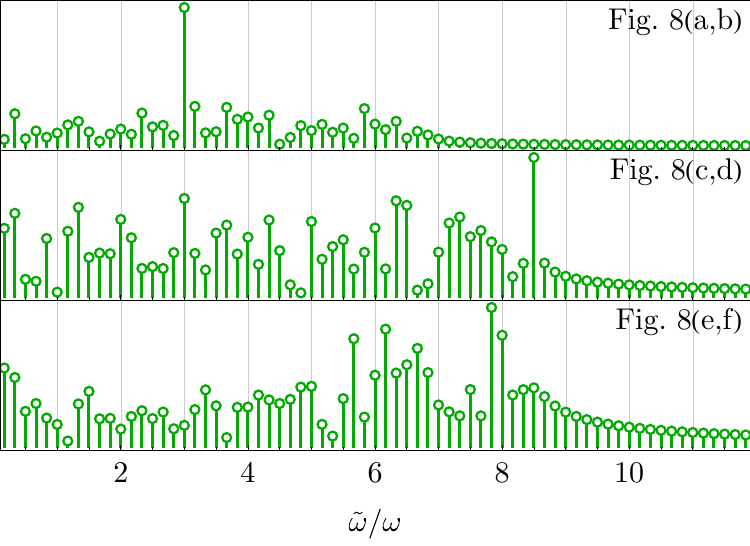}
    \caption{ Fourier transforms of displacement control functions $\PosControl(\Time)$ from Fig.~\ref{fig8}.
    The vertical axis is in arbitrary units, normalized to the highest frequency contribution.
    \label{FigD-8}}
\end{figure}

\begin{figure}[h!]
    \includegraphics[width=0.97\linewidth]{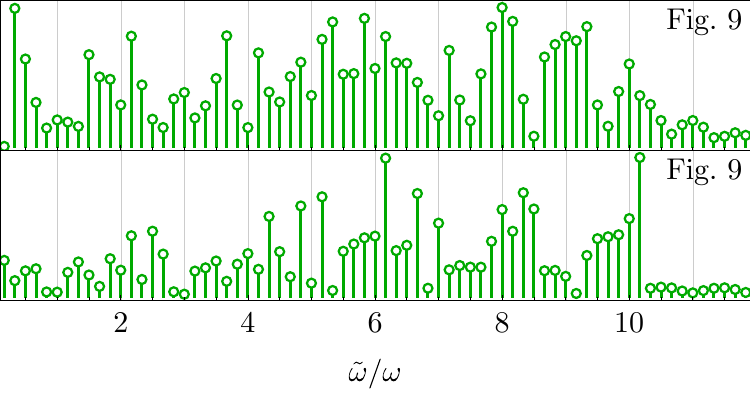}
    \caption{ Fourier transforms of displacement control functions $\PosControl(\Time)$ from Fig.~\ref{fig9} in the main text.
    The vertical axis is in arbitrary units, normalized to the highest frequency contribution.
    \label{FigD-9}}
\end{figure}

%\newpage

\renewcommand{\thefigure}{D-\arabic{figure}}
\setcounter{figure}{10}

\begin{figure}[h!]
    \includegraphics[width=\linewidth]{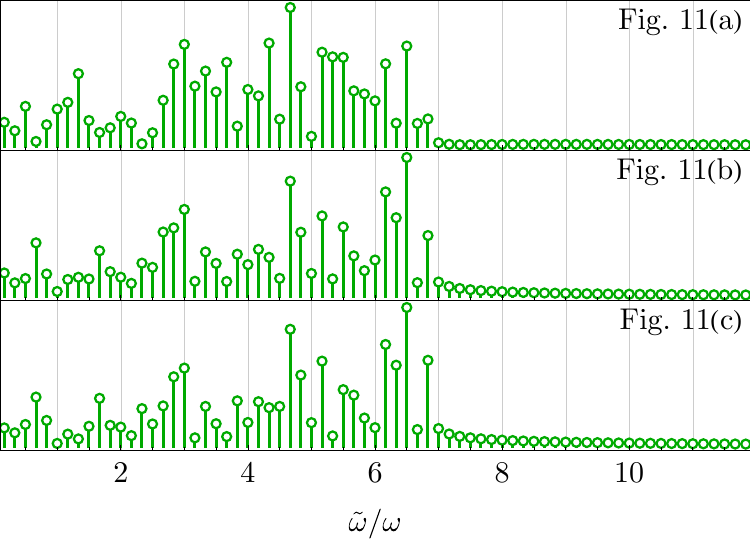}
    \caption{ Fourier transforms of displacement control functions $\PosControl(\Time)$ from Fig.~\ref{fig11}.
    The vertical axis is in arbitrary units, normalized to the highest frequency contribution.
    \label{FigD-11}}
\end{figure}

\renewcommand{\thefigure}{D-A\arabic{figure}}
\setcounter{figure}{0}

\begin{figure}[ht!]
    \includegraphics[width=\linewidth]{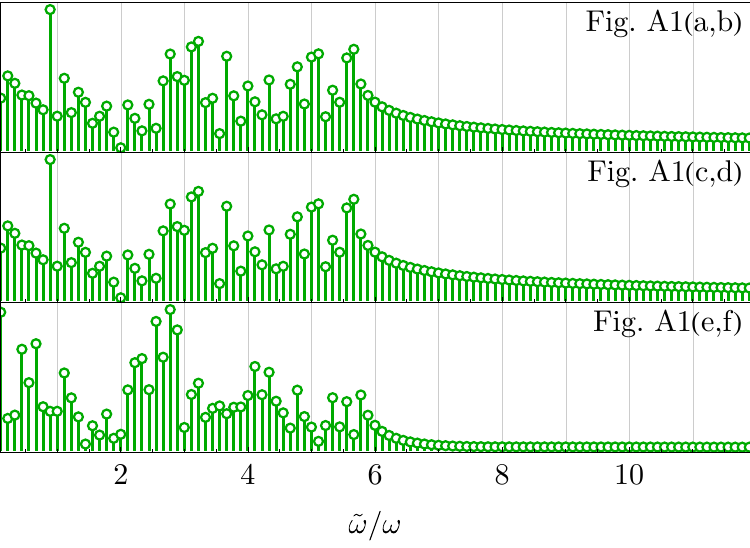}
    \caption{ Fourier transforms of displacement control functions $\PosControl(\Time)$ from Fig.~\ref{figA1}.
    The vertical axis is in arbitrary units, normalized to the highest frequency contribution.
    \label{FigD-A1}}
\end{figure}

\newpage
\renewcommand{\thefigure}{D-B\arabic{figure}}
\setcounter{figure}{0}

\begin{figure}[h!]
    \includegraphics[width=\linewidth]{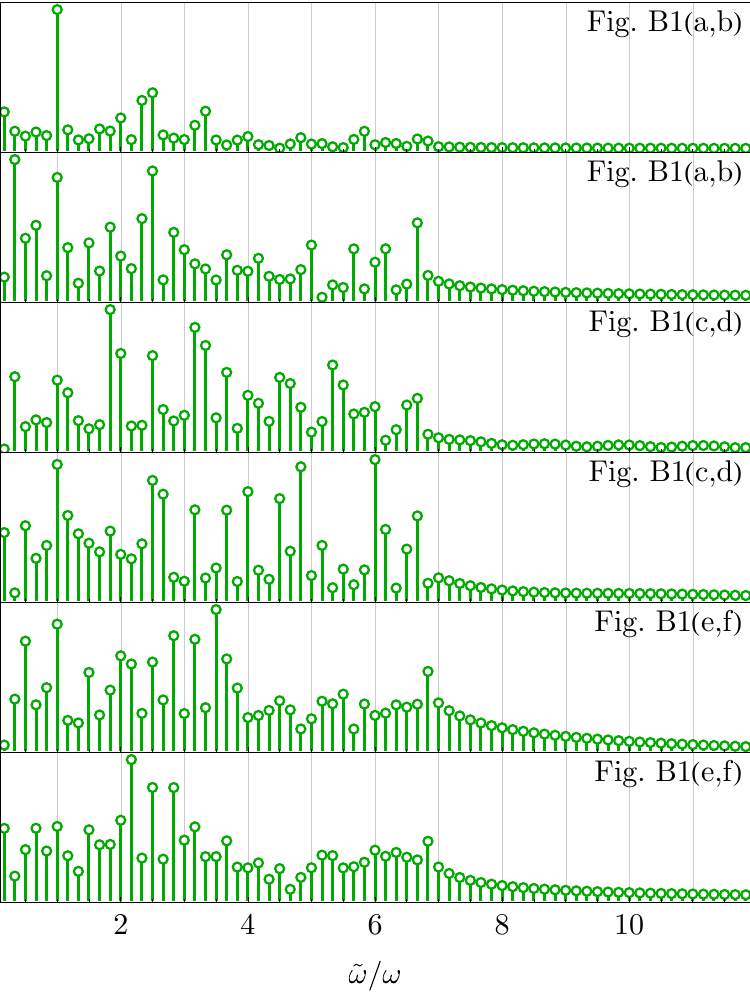}
    \caption{ Fourier transforms of displacement control functions $\PosControl(\Time)$ from Fig.~\ref{figB1}.
    The vertical axis is in arbitrary units, normalized to the highest frequency contribution.
    \label{FigD-B1}}
\end{figure}
\newpage

\renewcommand{\thefigure}{D-C\arabic{figure}}
\setcounter{figure}{0}

\begin{figure}[h!]
    \includegraphics[width=\linewidth]{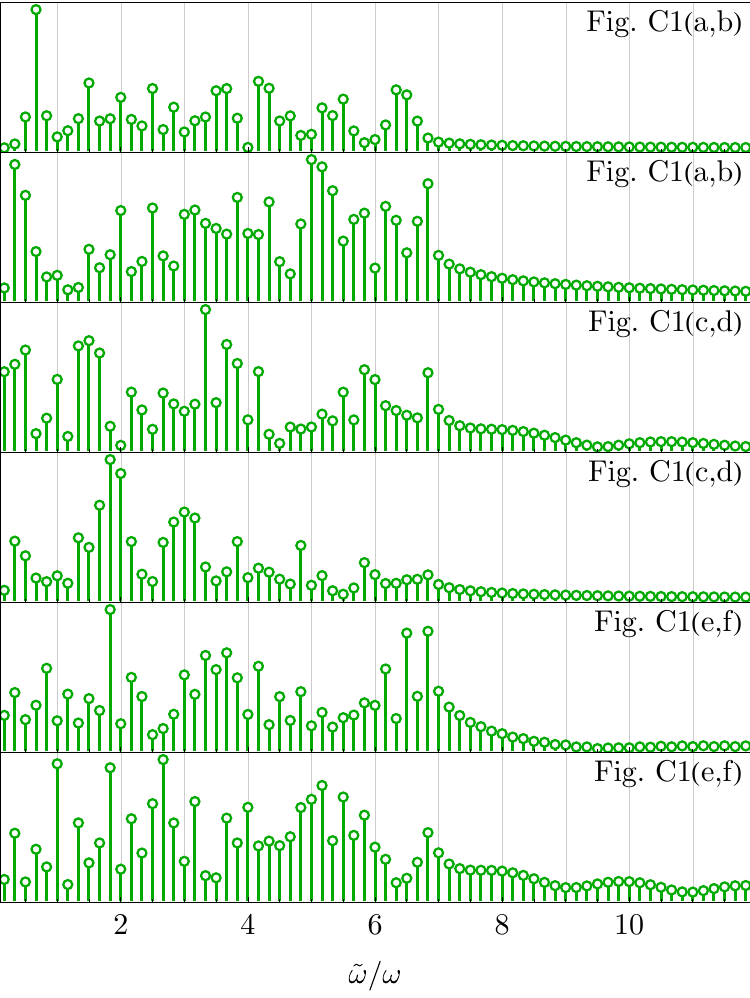}
    \caption{ Fourier transforms of displacement control functions $\PosControl(\Time)$ from Fig.~\ref{figC1}.
    The vertical axis is in arbitrary units, normalized to the highest frequency contribution.
    \label{FigD-C1}}
\end{figure}

\newpage

\section{Depth modulations}\label{depth}
Some of the protocols, presented in both the main text and the supplementary materials, utilized additional optimal control of the potential's depth, $\InControl(\Time)$.
In Figs.~\ref{FigE-5}, \ref{FigE-6}, and \ref{FigE-8}, these control functions are shown.

\renewcommand{\thefigure}{E-\arabic{figure}}
\setcounter{figure}{4}

\begin{figure}[h!]
    \includegraphics[width=\linewidth]{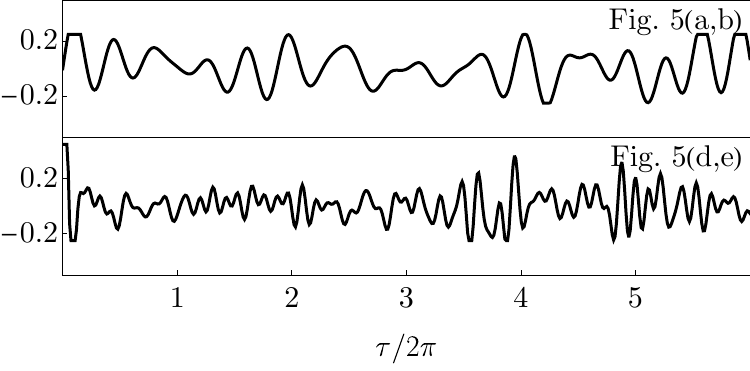}
    \caption{ Depth control functions $\InControl(\Time)$ from Fig.~\ref{fig5}.
    \label{FigE-5}}
\end{figure}

\begin{figure}[h!]
    \includegraphics[width=\linewidth]{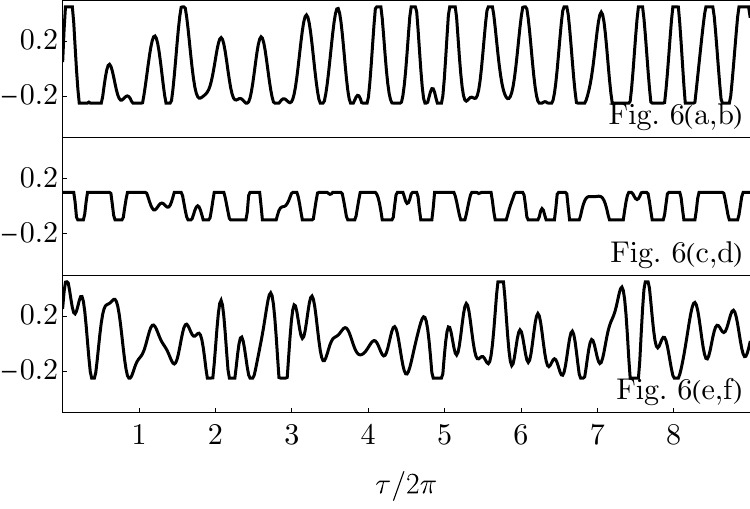}
    \caption{ Depth control functions $\InControl(\Time)$ from Fig.~\ref{fig6}.
    \label{FigE-6}}
\end{figure}

\renewcommand{\thefigure}{E-\arabic{figure}}
\setcounter{figure}{7}

\begin{figure}[h!]
    \includegraphics[width=\linewidth]{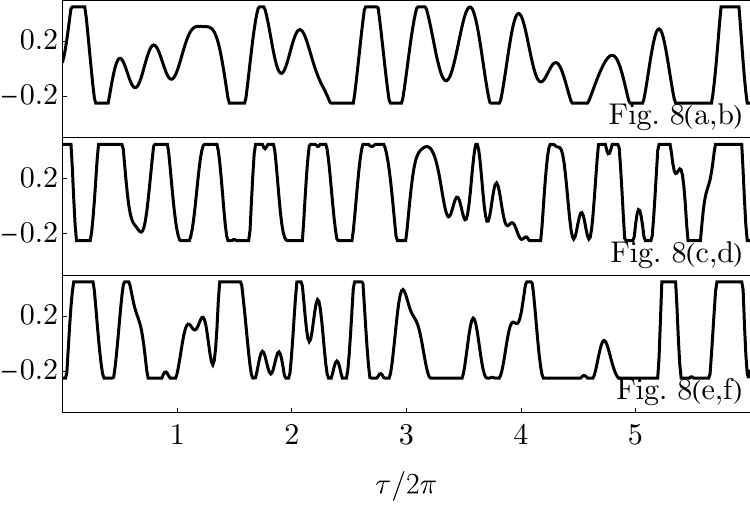}
    \caption{ Depth control functions $\InControl(\Time)$ from Fig.~\ref{fig8}.
    \label{FigE-8}}
\end{figure}

\end{document}